\documentclass[11pt]{article}
\usepackage{fullpage}
\usepackage{graphicx}
\usepackage{latexsym,amsmath,amssymb,amsthm,epic,eepic,multirow}
\usepackage{natbib}

\usepackage{multirow}
\usepackage{multicol}
\usepackage{subfigure} 
\usepackage{natbib}
\usepackage{algorithm}
\usepackage[noend]{algpseudocode}
\usepackage{mdframed}
\usepackage{tikz}
\usetikzlibrary{positioning}
\usepackage{pgf,tikz}
\usepackage{hyperref}
\hypersetup{ hidelinks }
\usepackage{listings} 
\usepackage{color} 
\usepackage{bbm}
\usepackage{relsize}
\usepackage{array}
\usepackage{mathtools}
\newcolumntype{P}[1]{>{\centering\arraybackslash}p{#1}}

\definecolor{upmaroon}{rgb}{0.48, 0.07, 0.07}
\definecolor{royalazure}{rgb}{0.0, 0.22, 0.66}
\definecolor{pakistangreen}{rgb}{0.0, 0.4, 0.0}

\newcommand{\PP}{\mathbb{P}}
\newcommand{\EE}{\mathbb{E}}

\theoremstyle{definition}
\newtheorem{theo}{Theorem}
\newtheorem{prop}{Proposition}
\newtheorem{lemm}{Lemma}

\newtheoremstyle{dotless}{}{}{}{}{\bfseries}{}{ }{}
\theoremstyle{dotless}
\newtheorem{assa}{}

\newtheorem{assap}{}

\newtheorem{assapp}{}

\newtheorem{assb}{}

\usepackage{color}

\let\originalleft\left
\let\originalright\right
\renewcommand{\left}{\mathopen{}\mathclose\bgroup\originalleft}
\renewcommand{\right}{\aftergroup\egroup\originalright}
\makeatletter
\newcommand{\leqnomode}{\tagsleft@true}
\newcommand{\reqnomode}{\tagsleft@false}
\makeatother

\begin{document}

\title{Assessing the overall and partial causal well-specification of nonlinear additive noise models}  
  
\author{Christoph Schultheiss and Peter B\"uhlmann\\
Seminar for Statistics, ETH Z\"urich}

\maketitle

\begin{abstract}
We propose a method to detect model misspecifications in nonlinear causal additive and potentially heteroscedastic noise models. We aim to identify predictor variables for which we can infer the causal effect even in cases of such misspecification. We develop a general framework based on knowledge of the multivariate observational data distribution. We then propose an algorithm for finite sample data, discuss its asymptotic properties, and illustrate its performance on simulated and real data.
\end{abstract}
\section{Introduction}
Nonlinear additive noise models and their heteroscedastic extensions are a popular modelling framework for causal discovery and inference. They allow to infer the true causal connections and effects from the multivariate distribution when the nonparametric model is correct; see, e.g., \cite{hoyer2008nonlinear,peters2014causal} or, for heteroscedastic models, \cite{strobl2022identifying, immer2022loci}. However, the conclusions can be misleading if the additive noise model is misspecified,
especially in the presence of hidden confounding variables. In this paper, we define the term  ``causal well-specification'' of additive noise models, discuss its relevance, and finally present a corresponding estimation technique for observational data.

The concept of well-specification for regression functionals in parametric regression was introduced by \cite{buja2019models2}. A regression functional is well-specified for a conditional target distribution if it only depends on the conditional distribution but is invariant to shifts in the predictors' distribution. This relates to the work by \cite{peters2016causal} and, hence, gives the notion of well-specification a causal interpretation. \cite{buja2019models2} suggest a set of reweighting diagnostics to assess well-specification of regression functions. For the linear model, an explicit test with asymptotic level as well as precise per-covariate interpretation for certain models is presented by \cite{schultheiss2023higher}.

If there is no functional assumption for the additive noise model, one must rely on flexible nonparametric regression techniques that approximate the conditional mean.
Considering well-specification of the conditional mean is of little use. It is by definition a property of the conditional distribution only. Hence, it is, upon existence, well-specified for arbitrary data generating mechanisms.

Thus, different concepts are needed to infer whether the estimated effects in an assumed additive noise model are causal. One of our contributions is the definition of causal well-specification and presenting its interpretation. Apart from global causal well-specification, we also define a local, i.e., per predictor version that is to be considered when the overall model does not satisfy the desired properties. This local viewpoint is of particular interest in the presence of hidden common causes, hidden mediating variables, or misspecified functional form, e.g., the effect of the unmeasured independent error cannot be separated as an individual addend. We propose a methodology to assess causal well-specification from observational data by relying on and exploiting conditional independence. Based on this, we derive an algorithm for finite sample data and prove its consistency. From a practical viewpoint, our estimated set of well-specified predictors (i.e., covariables) can be viewed as the one where the data is compatible (i.e., does not falsify) with the corresponding local structure of the model and its (partial) causal interpretation.

Almost no work exists on local goodness of fit or well-specification of nonlinear causal models, where local well-specification has a causal interpretation. The latter is the main goal of the present paper. Our method works in arbitrary structural causal models, i.e., if there is no well-specification, the interpretation becomes conservative but not wrong.

Causal structure learning with hidden variables in greatest generality is treated within the framework of Fast Causal Inference (FCI, \cite{spirtes2001anytime}). More specific and weakly related to our work is the approach by \cite{maeda2021causal} which discusses hidden variables in causal additive models (CAM, \cite{buhlmann2014cam}): unlike our current work, \cite{maeda2021causal} rely on the correctness of the causal additive model assumption. 
They present a causal graph detection algorithm based on unconditional independence tests.
We do not provide a graph search technique but consider verification or falsification of assumed causal structures instead.
We allow for as much flexibility in the model as possible while the CAM restricts the causal effect to sums of univariate functions. With the method in \cite{maeda2021causal} which is based on unconditional independence tests, certain edges remain undirected.
By considering conditional independence, additional edges could be directed - at least with a conditional independence oracle which is at the basis of our approach. After introducing our theory, we present an example in Section \ref{local} which illustrates some gains over the approach by \cite{maeda2021causal} by exploiting conditional independence.

\section{Causal well-specification in population}\label{sec:causal-wellspec}
We consider first the population case in which we know the joint distribution of the observed random variables, e.g. conditional expectations and conditional independence between random variables can be perfectly assessed. This section is a stand-alone and can be used in connection with other estimation algorithms than the ones presented in Sections \ref{estim} and \ref{subsec.asymp}.

In Section \ref{SCM}, we introduce the causal model, our notation, and the most important background concepts from the causality literature. Section \ref{roadmap} provides a ``roadmap'' of our methodology. We describe on a high level which assumptions lead to a causal interpretation of the additive noise model, and how we can assess these using conditional independence statements. The mathematical details around these concepts follow in Sections \ref{global} and \ref{local}.

\subsection{Structural causal model}\label{SCM}
We summarize the concepts from the causality literature that are fundamental to our work.
Let $\mathbf{Z} = \left(Z_1, \ldots, Z_q\right) ^\top \in \mathbb{R}^q$ be a random vector whose entries $Z_j$ follow a structural causal model (SCM), say, $\mathfrak{C}$,
\begin{equation}\label{eq:SEM}
\mathfrak{C}: \quad Z_j \leftarrow f_j\left(\mathbf{Z}_{\text{PA}\left(j\right)}, \xi_j \right) \ \forall j \in \left\{1,\ldots,q\right\}.
\end{equation}
We write $\leftarrow$ to emphasize that the equality is induced by a causal effect. $\xi_j$ is some noise that is jointly independent over $j$. The set $\text{PA}\left(j\right)$ denotes the parents of $j$, i.e., covariates $Z_k$ with $k \in \text{PA}\left(j\right)$ have a direct causal effect on $Z_j$. Conversely, $j$ is a child of $k$. The SCM is represented by a directed graph that has an edge from $Z_k \rightarrow Z_j$ if and only if $k \in \text{PA}\left(j\right)$. We assume that this results in a directed acyclic graph (DAG).

If the DAG contains any directed path from $k$ to $j$, which may include several edges, we call $j$ a descendant of $k$, $j \in \text{DE}\left(k\right)$, and $k$ an ancestor of $j$. On a path, $Z_k \rightarrow Z_l \rightarrow Z_j$, we call $l$ a mediator. In a structure $Z_k \leftarrow Z_l \rightarrow Z_j$, $l$ is a confounder.

We use the concept of d-separation \citep[Section~3]{geiger1990identifying}. Two sets of variables $\mathbf{Z}_A$ and $\mathbf{Z}_B$ are d-separated by $\mathbf{Z}_C$ if it blocks all paths from $\mathbf{Z}_A$ to $\mathbf{Z}_B$. There are two ways to block a path:
\begin{itemize}
\item $\exists j \in C$ such that $Z_k \rightarrow Z_j \rightarrow Z_l$, $Z_k \leftarrow Z_j \leftarrow Z_l$, or $Z_k \leftarrow Z_j \rightarrow Z_l$ is on the path.
\item $\exists j \not\in C$ such that $Z_k \rightarrow Z_j \leftarrow Z_l$ is on the path, and $\left(\text{DE}\left(j\right) \cap C\right) = \emptyset$.
\end{itemize}
The joint independence of the $\xi_j$ in \eqref{eq:SEM}, implies that d-separated sets of variables are independent conditioned on the separating set \citep[Theorem 1.4.1]{pearl2009causality}. This is called the global Markov property. It applies for unconditional independence with $C=\emptyset$ as well. The distribution is called faithful to its DAG if all independences are implied by such a d-separation \citep[Chapter~2.3.3]{spirtes2000causation}. Violations of faithfulness can intuitively be described as cancellations of effects such that dependencies that one would assume to exist from the graph alone vanish.

We are interested in the situation where one variable with index in $\left\{1,\ldots,q\right\}$ is the target, some of the variables are observed (potential) predictors and the rest are unobserved or ignored (potential) predictors. Let $Y$, $M$ (\textbf{m}easured) and $N$ (\textbf{n}ot measured) be a partition of $\left\{1,\ldots,q\right\}$ that represent these subsets, and define the corresponding random variables and vectors
\begin{equation*}
Y \coloneqq Z_Y, \quad \mathbf{X} \coloneqq \mathbf{Z}_M \in \mathbb{R}^p, \quad  \mathbf{H} \coloneqq \mathbf{Z}_N, \quad \mathbf{X}_{\text{PA}\left(Y\right)} \coloneqq \mathbf{Z}_{M \cap \text{PA}\left(Y\right)} \quad \text{and} \quad \mathbf{H}_{\text{PA}\left(Y\right)} \coloneqq \mathbf{Z}_{N \cap \text{PA}\left(Y\right)}.
\end{equation*}
In words, $Y$ is the target, $\mathbf{X}$ are observed covariates, $\mathbf{H}$ are latent variables, $\mathbf{X}_{\text{PA}\left(Y\right)}$ is the subset of $Y$'s parents that we observe, and $\mathbf{H}_{\text{PA}\left(Y\right)}$ is the subset that we do not observe. With a slight abuse of notation, $Y$ can represent the target random variable or the index in $\left\{1,\ldots,q\right\}$ that corresponds to the target. Note that for notational simplicity, we can absorb $\xi_Y$ to be an additional variable in $\mathbf{H}_{\text{PA}\left(Y\right)}$. Therefore, $\mathbf{H}_{\text{PA}\left(Y\right)}$ always has dimensionality of at least one assuming $Y$ is not deterministic in $\mathbf{X}$. In our SCM \eqref{eq:SEM}, we then have
\begin{equation*}
Y \leftarrow f_Y\left(\mathbf{X}_{\text{PA}\left(Y\right)}, \mathbf{H}_{\text{PA}\left(Y\right)}\right).
\end{equation*}
For a realization $\mathbf{z}$ of $\mathbf{Z}$, we use the same naming convention, e.g., the realization of $\mathbf{X}$ is then $\mathbf{x}$.

We define the term Markov blanket. Consider $\mathbf{H}_{\text{PA}\left(Y\right)}$ as an exemplary target, analogous definitions for other targets exist.  We call a set $S$ a Markov blanket of these hidden parents if
\begin{equation*}
\mathbf{H}_{\text{PA}\left(Y\right)} \perp \mathbf{X}_{-S} \vert \mathbf{X}_S,
\end{equation*}
where $\mathbf{X}_{-S}$ denotes all observed variables that are not in $S$. Importantly, we always mean these blankets to be found within only the observed covariates $\mathbf{X}$. Markov blankets are also known as sufficient sets. We define minimal Markov blankets as Markov boundaries, i.e., a set $S$ such that
\begin{equation*}
\mathbf{H}_{\text{PA}\left(Y\right)} \perp \mathbf{X}_{-S} \vert \mathbf{X}_S, \ \text{but} \ \forall S' \subset S: \mathbf{H}_{\text{PA}\left(Y\right)} \not \perp \mathbf{X}_{-S'} \vert \mathbf{X}_{S'}.
\end{equation*}
As the Markov boundary is defined within only the observed covariates $\mathbf{X}$, in a structure $H_1 \rightarrow H_2 \rightarrow X_1$, $X_1$ would still count as part of the boundary of $H_1$ since it is the nearest measured descendant.
We discuss the uniqueness of the Markov boundary in Section \ref{local}. It could also be all of $\mathbf{X}$ or empty.

We use causal do-notation to denote interventions. Conditioning on, e.g., $\text{do}\left(Z_j \leftarrow z_j\right)$ means that we assume a variation of the SCM $\mathfrak{C}$ \eqref{eq:SEM} where the structural assignment for $Z_j$ is not a function in its parents and the noise term but set to a fixed value. The remaining structural equations remain the same. Similarly, $\text{do}\left(\mathbf{Z}_S \leftarrow \mathbf{z}_S\right)$ means that a whole set of variables is intervened to have a fixed value.

We also apply the related concept of counterfactuals: what would happen to an observed data point if some of the covariates are set to hard values while the remaining structural assignments and unobserved noise terms remain unaffected? We use the notation from \cite[Chapter~6.4]{peters2017elements}, i.e., $P_Y^{\mathfrak{C} \vert \mathbf{Z}=\mathbf{z}; \text{do}\left(\mathbf{X}\leftarrow \mathbf{x}'\right)}$ denotes the counterfactual distribution of $Y$ in the SCM $\mathfrak{C}$ where $\mathbf{Z}=\mathbf{z}$ is observed and the counterfactual intervention is $\mathbf{X}\leftarrow \mathbf{x}'$.

\subsection{Roadmap of our methodology}\label{roadmap}
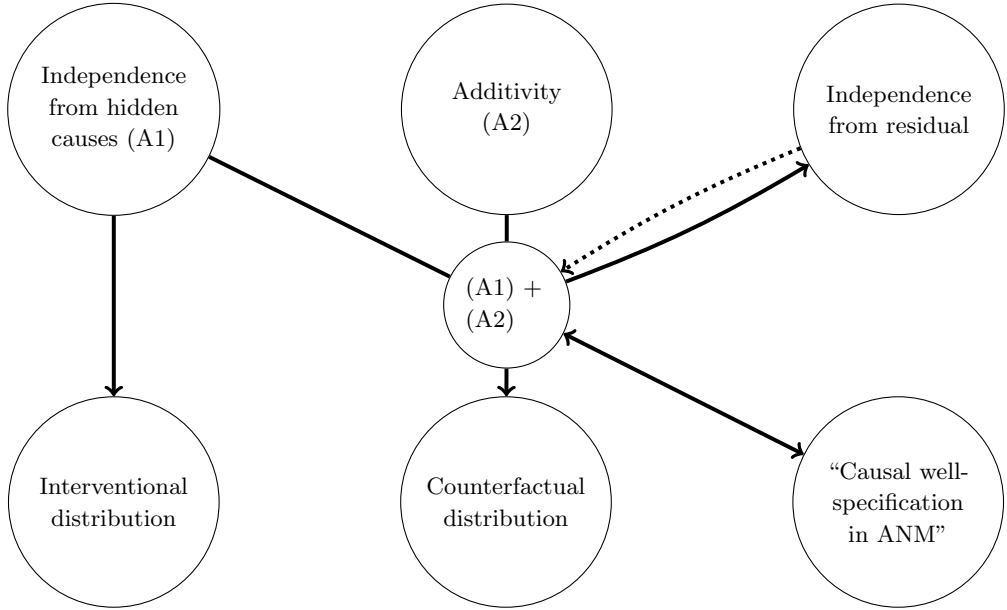
\begin{figure}[h!]
\centering
    \begin{tikzpicture}[node distance=2.4cm, every node/.style={font=\fontsize{9}{12}\selectfont}]
        \node (IH) [draw, circle, minimum size=2.8cm, text width=2.2cm, align=center] {Independence from hidden causes (A1)};
        \node (Ad) [draw, circle, minimum size=2.8cm, text width=2.2cm, align=center, right=of IH] {Additivity (A2)};
        \node (plus) [draw, circle, minimum size=0.5cm, text width = 1.1cm, below= 0.35cm of Ad] {(A1) + (A2)};
        \node (ID) [draw, circle, minimum size=2.8cm, text width=2.2cm, align=center, below = of IH] {Interventional distribution};
        \node (CD) [draw, circle, minimum size=2.8cm, text width=2.2cm, align=center, right = of ID] {Counterfactual distribution};
        \node (IR) [draw, circle, minimum size=2.8cm, text width=2.2cm, align=center, right = of Ad] {Independence from residual};
        \node (CW) [draw, circle, minimum size=2.8cm, text width=2.2cm, align=center, right = of CD] {``Causal well-specification in ANM''};
        
        \draw[->, line width=1.5pt] (IH) -- (ID);
        \draw[-, line width=1.5pt] (IH) -- (plus);
        \draw[-, line width=1.5pt] (Ad) -- (plus);
        \draw[->, line width=1.5pt] (plus) -- (CD);
        \draw[<->, line width=1.5pt] (plus) to (CW);
        \draw[->, line width=1.5pt, bend right=5] (plus) to (IR);
        \draw[dotted, ->, line width=1.5pt, bend right=5] (IR) to (plus);
    \end{tikzpicture}
 \caption[Interconnections]
 {Interconnection between assumptions, causal interpretations, and the testable proxy. Directed edges denote implications, the plus means that both assumptions hold simultaneously, the bidirected edge denotes equivalence per definition, and the dotted edge denotes implication up to pathological cases, i.e., a proxy.}
 \label{fig:roadmap}
\end{figure}

We describe here on a high level the idea of causal well-specification of the general additive noise model as defined below. The interplay between the different assumptions, their causal implications, and how we aim to test for it is then visualized in Figure \ref{fig:roadmap}. Detailed mathematical definitions, assumptions, and results are given in the subsequent sections.

The additive noise model (ANM) has the following structure
\begin{equation*}
Y \leftarrow f_{\mathbf{X}Y}\left(\mathbf{X}_{\text{PA}\left(Y\right)}\right) + f_{\mathbf{H}Y}\left(\mathbf{H}_{\text{PA}\left(Y\right)}\right), \quad \text{where} \quad \mathbf{X} \perp \mathbf{H}_{\text{PA}\left(Y\right)}.
\end{equation*}
It implies a testable proxy
\begin{equation*}
H_0: \quad  \mathcal{E} \perp \mathbf{X}, \quad \text{where} \quad \mathcal{E} = Y - \EE\left[Y\vert \mathbf{X} \right].
\end{equation*}
Independence from the hidden causes means that for the outcome's distribution it makes no difference if we observe $\mathbf{X}=\mathbf{x}$ or enforce it by intervention. With the additivity, we can even understand how the outcome reacts to a counterfactual change in $\mathbf{X}$ while keeping the unobserved $\xi_k$ in \eqref{eq:SEM}  $\forall k \in N$ fixed. Hence, we can understand how the system reacts to change purely from the observational distribution.

The ANM assumption or the null-hypothesis $H_0$ can be violated in presence of dependence between $\mathbf{X}$ and hidden causal parents of $Y$, or misspecified functional form, meaning interactions between $\mathbf{X}_{\text{PA}\left(Y\right)}$ and the hidden parents of $Y$ in the structural equation for $Y$, or both. However, it can be that some observed covariates do not interact with the hidden causes (\ref{ass:add-sep} later) and are conditionally independent of these given the remaining covariates (\ref{ass:markov} later). We then say the ANM is causally well-specified for these covariates, say, $\mathbf{X}_U$. This implies another testable proxy
\begin{equation*}
H_{0,U}: \quad \mathcal{E} \perp \mathbf{X}_U \vert \mathbf{X}_{-U}, \quad \text{with} \quad \mathcal{E} = Y - \EE\left[Y\vert \mathbf{X} \right].
\end{equation*}
As for all observed variables, $M$, we do not restrict $U$ to be among the parents of $Y$ since we typically do not have knowledge of these. Our causal interpretation of well-specification remains valid even for $j \in U \setminus \text{PA}\left(Y\right)$: we can correctly characterize the absence of effects.
Conditional independence from the unobserved causes means that for the outcome's distribution it makes no difference if we observe $\mathbf{X}_U=\mathbf{x}_U$ or enforce it by intervention at fixed levels of $\mathbf{X}_{-U}$. With the additivity, we can even understand how the outcome reacts to a counterfactual change in $\mathbf{X}_U$ while keeping $\mathbf{X}_{-U}$ and the unobserved $\xi_k$ in \eqref{eq:SEM}  $\forall k \in N$ fixed. Hence, we can understand how the system reacts to changes in some covariates purely from the observational distribution.

By definition, such $\left\{1, \ldots, p \right\} \setminus U$ is a Markov blanket of $\mathcal{E}$. The more variables we can put into $U$ the more explicative our model becomes. Hence, we aim to find a Markov boundary. But, having any blanket is enough to avoid false causal claims. We summarize in Figure \ref{fig:roadmap}.

Importantly, our results do not assume any model apart from the SCM in \eqref{eq:SEM}. We require independence and additivity to identify causal implications of the ANM. But, we can correctly falsify models if these assumptions do not hold.

Such causal well-specification can also be of use if one is mainly interested in purely predictive tasks and aims for out-of-distribution generalization where the new (test) data distribution is different (``shifted'') from the training distribution; see, e.g., \cite{rojas2018invariant} or the survey by \cite{wang2022generalizing}. Consider the task of domain adaptation with only few data in a target domain but many observations in a different training domain. If the distribution of the $\xi_j$ in the target environment is shifted, also the best predictive function, $\EE\left[Y \vert \mathbf{X}\right]$ might be different such that the large training set is not suitable for learning the predictive function. But, assuming invariant causal assignments in our SCM \eqref{eq:SEM}, the addend of the conditional mean induced by the causally well-specified covariates remains invariant across such shifted domains. Hence, this invariant part of the function could be estimated using the large multi-source data set from different domains.

\subsection{Global well-specification}\label{global}
We recapitulate the ANM assumption for covariates $\mathbf{X}$ and target $Y$. We call the ANM causally well-specified if
\begin{equation}\label{eq:ANM}
Y \leftarrow f_{\mathbf{X}Y}\left(\mathbf{X}_{\text{PA}\left(Y\right)}\right) + f_{\mathbf{H}Y}\left(\mathbf{H}_{\text{PA}\left(Y\right)}\right), \quad \text{where} \quad \mathbf{X} \perp \mathbf{H}_{\text{PA}\left(Y\right)}.
\end{equation}
Note that we do not constraint the functional form of $f_{\mathbf{X}Y}\left(\cdot\right)$ and $f_{\mathbf{H}Y}\left(\cdot\right)$ any further. In particular, the structure does not imply that the functions must be additive in their respective arguments.

The independence condition corresponds to no hidden confounding or hidden mediation. It ensures 
\begin{equation*}
Y\vert \mathbf{X}=\mathbf{x} \quad \overset{d}{=} \quad Y\vert \text{do}\left(\mathbf{X}\leftarrow \mathbf{x}\right),
\end{equation*}
where $\overset{d}{=}$ states that two random variables have the same distribution.
Assuming faithfulness, it also implies $\text{DE}\left(Y\right) \cap M = \emptyset$ since faithfulness ensures  $\forall j \in \text{DE}\left(Y\right)\ Z_j \not \perp \xi_Y \in \mathbf{H}_{\text{PA}\left(Y\right)}$.

The parametrization in \eqref{eq:ANM} is not unique as constants could be moved between the two summands. We let the second have mean $0$ such that $\EE\left[Y\vert \mathbf{X} \right]  = f_{\mathbf{X} Y}\left(\mathbf{X}_{\text{PA}\left(Y\right)}\right)$. The additivity condition then ensures that in the counterfactual, where we can change $\mathbf{X}$ without changing any other unobserved noise term, the outcome is exactly shifted by the difference in conditional expectation. Thus, we fully understand the effect of changing $\mathbf{X}$.
Denote point masses at $y$ by $\delta_y$, then,
\begin{equation*}
P_Y^{\mathfrak{C} \vert \mathbf{Z}=\mathbf{z}; \text{do}\left(\mathbf{X}\leftarrow \mathbf{x}'\right)}=\delta_{y'} \quad \text{where} \quad y'=y + \EE\left[Y\vert \mathbf{X} = \mathbf{x}'\right]-\EE\left[Y\vert \mathbf{X} = \mathbf{x}\right].
\end{equation*}

The conditions in \eqref{eq:ANM} additionally imply the following global null hypothesis that we aim to check first.
\begin{equation}\label{eq:H0}
H_0: \quad  \mathcal{E} \perp \mathbf{X}, \quad \text{where} \quad \mathcal{E} = Y - \EE\left[Y\vert \mathbf{X} \right].
\end{equation}
Note the subtle difference between \eqref{eq:ANM} and \eqref{eq:H0}.
The conditions in \eqref{eq:ANM} are sufficient to fulfill \eqref{eq:H0} but they are not necessary as such independence could also exist non-causally. A prime example is with jointly Gaussian $\mathbf{Z}$: then, $H_0$ holds regardless of the independence condition in \eqref{eq:ANM} as $\mathcal{E}$ and $\mathbf{X}$ are uncorrelated. The additivity is always fulfilled since multivariate Gaussianity implies linear additive causal effects. However, except for Gaussian $\mathbf{Z}$ or some other pathological data generating distributions, \eqref{eq:H0} is a useful proxy for \eqref{eq:ANM}, i.e., it allows to check whether $\EE\left[Y\vert \mathbf{X} \right]$ represents a true causal effect; see also the discussions on the identifiability of ANM in \cite{hoyer2008nonlinear} and \cite{peters2014causal}.

To test $H_0$, any valid test for independence of $\mathbf{X}$ and $\mathcal{E}$ can be used.

\subsection{Local well-specification}\label{local}
\begin{figure}[h!]
\centering
\begin{tikzpicture}[roundnode/.style={circle, very thick, minimum size=8mm}]
\node[draw, roundnode, text centered] (x1) {${\scriptstyle X_1}$};
\node[draw, roundnode, right =of x1, text centered] (x2) {${\scriptstyle X_2}$};
\node[draw, roundnode, right =of x2, text centered] (y) {${\scriptstyle Y}$};
\node[draw, dotted, roundnode, above =of x2, text centered] (h) {${\scriptstyle H}$};

\draw[->, line width= 1] (x1) -- (x2);
\draw[->, line width= 1] (x2) -- (y);
\draw[->, line width= 1] (h) -- (x1);
\draw[->, line width= 1] (h) -- (y);

\node[draw, roundnode, right = of y] (x2b) {${\scriptstyle X_2}$};
\node[draw, dotted, roundnode, right =of x2b, text centered] (hb) {${\scriptstyle H}$};
\node[draw, roundnode, right =of hb, text centered] (yb) {${\scriptstyle Y}$};
\node[draw, roundnode, above =of hb, text centered] (x1b) {${\scriptstyle X_1}$};

\draw[->, line width= 1] (x1b) -- (x2b);
\draw[->, line width= 1] (x2b) -- (hb);
\draw[->, line width= 1] (x1b) -- (yb);
\draw[->, line width= 1] (hb) -- (yb);
\end{tikzpicture}
 \caption[Sample DAGs]
 {Left: Structure with a hidden confounder. Right: Structure with a hidden mediator.}
 \label{fig:DAGs}
\end{figure}
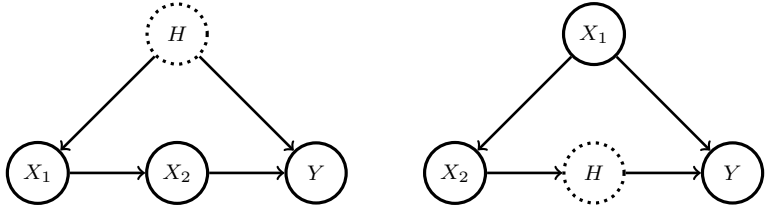

If the conditions \eqref{eq:ANM} are partially violated it might still be possible to correctly understand the causal effect for \emph{some} of the predictors $\mathbf{X}_U$ where $U \subseteq \left\{1, \ldots, p \right\}$. We say the effect of $\mathbf{X}_U$ is causally well-specified in the ANM with response $Y$ and covariates $\mathbf{X}$ if the following hold.
\begin{assa}\label{ass:markov}
The covariates in $\mathbf{X}_{-U}$ form a Markov blanket of $\mathbf{H}_{\text{PA}\left(Y\right)}$, i.e., $\mathbf{H}_{\text{PA}\left(Y\right)} \perp \mathbf{X}_U \vert \mathbf{X}_{-U}$.
\end{assa}
\begin{assa}\label{ass:add-sep}
$Y \leftarrow f_{\mathbf{X}_U Y}\left(\mathbf{X}_U, \mathbf{X}_{\text{PA}\left(Y\right) \setminus U}\right) + f_{\mathbf{H}Y}\left(\mathbf{H}_{\text{PA}\left(Y\right)}, \mathbf{X}_{\text{PA}\left(Y\right) \setminus U}\right)$, i.e., the causal effect is additively separable into all terms that include $\mathbf{X}_U$ only with observed $\mathbf{X}$ and all terms that include $\mathbf{H}$  without $\mathbf{X}_U$. 
\end{assa}
Consider Figure \ref{fig:DAGs} containing two examples of DAGs with hidden parents. On the left, the set $U=\left\{2\right\}$ fulfils \ref{ass:markov}. On the right, $U=\left\{1\right\}$ fulfils it. Correctness of \ref{ass:add-sep} depends on the structural assignment for $Y$. E.g.,
\begin{equation*}
Y \leftarrow \sin\left(X_2\right) + \sin\left(H\right),
\end{equation*}
is ok for $U=\left\{2\right\}$ in the example on the left, but
\begin{equation*}
Y \leftarrow \sin\left(X_2 + H\right),
\end{equation*}
is not. 

If $\exists k\in M \cap \text{DE}\left(Y\right)$, i.e., we observe one or more descendants of $Y$, all $X_j$ where $j \in \text{PA}\left(Y\right)$ are, up to faithfulness, in any Markov blanket of $\xi_Y \in \mathbf{H}_{\text{PA}\left(Y\right)}$ since they have a common child with respect to the measured covariates. Thus, sets containing measured parents do not fulfill \ref{ass:markov}. When choosing (causal) predictors, one aims for $M \cap \text{DE}\left(Y\right) = \emptyset$, but there is in general no guarantee for it. So, this is another potential model misspecification. Violations of faithfulness are irrelevant for \ref{ass:markov}: if there are effects from $\mathbf{H}_{\text{PA}\left(Y\right)}$ to $\mathbf{X}_U$ or vice-versa that cancel out each other, we receive the same implications as if there were no such effects unless \ref{ass:indep} is violated; see below.

We impose no constraints onto $f_{\mathbf{X}_U Y}\left(\cdot\right)$ and $f_{\mathbf{H} Y}\left(\cdot\right)$ in \ref{ass:add-sep}. Hence, the first summand could be zero, which is the case for $U \cap \text{PA}\left(Y\right) = \emptyset$.

\ref{ass:markov} ensures that
\begin{equation*}
Y\vert \mathbf{X}_U=\mathbf{x}_U, \mathbf{X}_{-U}=\mathbf{x}_{-U} \quad \overset{d}{=} \quad Y\vert \text{do}\left(\mathbf{X}_U \leftarrow \mathbf{x}_U\right), \mathbf{X}_{-U}=\mathbf{x}_{-U}
\end{equation*}
whenever both are defined. This follows from the second rule of do-calculus \citep{pearl2012calculus}. After removing edges out of $\mathbf{X}_U$, dependence between $\mathbf{X}_U$ and $Y$ could only be induced by a common ancestor or a path from $Y$ to $\mathbf{X}_U$. But, these are all blocked by $\mathbf{X}_{-U}$, on which we condition, by the assumption.

Combined with \ref{ass:add-sep}, we get two implications under an additional technical assumption.
\begin{assb}\label{ass:indep}
Let $\left\{A,B,C\right\}$ be disjoint subsets of $\left\{1,\ldots,q\right\}$ in model \eqref{eq:SEM}. Then,
\begin{equation*}
\mathbf{Z}_A \perp \mathbf{Z}_B \vert \mathbf{Z}_C \implies \mathbf{Z}_A  \vert \mathbf{Z}_B=\mathbf{z}_B, \mathbf{Z}_C=\mathbf{z}_C \ \overset{d}{=} \ \mathbf{Z}_A \vert \mathbf{Z}_B=\mathbf{z}'_B, \mathbf{Z}_C=\mathbf{z}_C \quad \forall \mathbf{z}_B, \mathbf{z}'_B, \mathbf{z}_C.
\end{equation*}
\end{assb}
This means that there are no unobservable dependencies on null sets of the observational distribution which is natural to assume except for pathological data. In general, independence only implies the latter equality for almost all $\mathbf{z}_B, \mathbf{z}'_B, \mathbf{z}_C$. A counterexample to \ref{ass:indep} would be the following SCM with continuous and univariate components $Z_A$, $Z_B$, and $Z_C$
\begin{align*}
Z_C & \leftarrow \xi_C, \\
Z_B & \leftarrow Z_C + \xi_B, \\
Z_A & \leftarrow Z_C + \xi_A + \mathbbm{1}_{\left\{Z_B = 0\right\}}.
\end{align*}
Then, 
\begin{equation*}
Z_A \perp Z_B \vert Z_C, \quad \text{but, e.g.,} \quad Z_A  \vert Z_B=1, Z_C=0 \ \overset{d}{\neq} \ Z_A \vert Z_B=0, Z_C=0,
\end{equation*}
i.e., there is no observable conditional dependence between $Z_A$ and $Z_B$, but we could, in theory, create an intervention that provokes unexpected behaviour. Therefore, we exclude such hidden dependencies.
\begin{theo}\label{theo:H0j}
Assume the model \eqref{eq:SEM} with \ref{ass:indep}. Let $\mathbf{X}_U$ be a set of covariates fulfilling \ref{ass:markov} and \ref{ass:add-sep}, then
\begin{equation*}
P_Y^{\mathfrak{C} \vert \mathbf{Z}=\mathbf{z}; \text{do}\left(\mathbf{X}_U \leftarrow \mathbf{x}'_U, \mathbf{X}_{-U} \leftarrow \mathbf{x}_{-U}\right)}=\delta_{y'} \quad \text{where} \quad y'=y + \EE\left[Y\vert \mathbf{X}_U = \mathbf{x}'_U, \mathbf{X}_{-U}=\mathbf{x}_{-U}\right]-\EE\left[Y\vert \mathbf{X} = \mathbf{x}\right]
\end{equation*}
for $ \left(\mathbf{X}_U = \mathbf{x}'_U, \mathbf{X}_{-U}=\mathbf{x}_{-U}\right)$ in the support of the observational distribution. Further, $H_{0,U}$ holds, where
\begin{equation}\label{eq:H0j}
H_{0,U}: \quad \mathcal{E} \perp \mathbf{X}_U \vert \mathbf{X}_{-U}, \quad \text{with} \quad \mathcal{E} = Y - \EE\left[Y\vert \mathbf{X} \right].
\end{equation}
\end{theo}

The first implication means that in the counterfactual, where we can change $\mathbf{X}_U$ without changing $\mathbf{X}_{-U}$ or $\xi_k$ in \eqref{eq:SEM} $\forall k \not \in M$, the effect on $Y$ is fully determined by the shift in conditional expectation. Thus, we understand the causal effect of this theoretical intervention. Note that with \ref{ass:markov} and \ref{ass:indep}, not changing $\mathbf{X}_{-U}$ and $\xi_k \ \forall k \not \in M$ is equivalent to not changing $\mathbf{X}_{-U}$ and $\mathbf{H}$, i.e., all other variables apart from $\mathbf{X}_U$ and $Y$ remain unchanged; see also the proof in Appendix \ref{proof:H0j}. More generally, including cases where $ \left(\mathbf{X}_U = \mathbf{x}'_U, \mathbf{X}_{-U}=\mathbf{x}_{-U}\right)$ is outside the support of the observational distribution, one could replace
\begin{equation*}
\EE\left[Y\vert \mathbf{X}_U = \mathbf{x}'_U, \mathbf{X}_{-U}=\mathbf{x}_{-U}\right] \quad \text{by} \quad \EE\left[Y\vert \text{do}\left(\mathbf{X}_U \leftarrow \mathbf{x}'_U\right), \mathbf{X}_{-U}=\mathbf{x}_{-U}\right]
\end{equation*}
which are equivalent if both are defined as discussed above.
However, this is not estimable outside the data support. $H_{0,U}$ \eqref{eq:H0j} is equivalent to saying that $\left\{1, \ldots p\right\} \setminus U$ defines a Markov blanket of the residual $\mathcal{E}$.

We note that the implication of \ref{ass:markov} would be of practical interest on its own. However, as we do not know of any useful proxy for it that can be calculated by the observational distribution, we always consider the combination of \ref{ass:markov} and \ref{ass:add-sep} as the object of interest.

The local null hypothesis $H_{0,U}$, which can be checked by the observational distribution alone, serves as a proxy for \ref{ass:markov} and \ref{ass:add-sep}. Again, a multivariate Gaussian distribution is an example where \eqref{eq:H0j} holds regardless of \ref{ass:markov}. However, for other data generating distributions, we consider \eqref{eq:H0j} to be a good proxy to see whether \ref{ass:markov} and \ref{ass:add-sep} might hold.

Of most interest are the sets
\begin{equation}\label{eq:mb}
W \in \underset{U: H_{0,U} \ \text{is true}}{\text{arg max}}\left\vert U \right\vert.
\end{equation}
As $W$ is of maximum size, $\left\{1, \ldots p\right\} \setminus W$ is of minimum size. Thus, it is not only a Markov blanket but a Markov boundary of $\mathcal{E}$ such that uniqueness of $W$ is implied by the uniqueness of the Markov boundary. This is guaranteed if the so-called intersection property holds \citep[Chapter~3]{pearl1988probabilistic}.
\begin{assb}\label{ass:inter}
$\mathcal{E} \perp \mathbf{X}_A \vert \mathbf{X}_B, \mathbf{X}_C \quad \text{and} \quad \mathcal{E} \perp \mathbf{X}_B  \vert \mathbf{X}_A, \mathbf{X}_C \implies \mathcal{E} \perp \mathbf{X}_A, \mathbf{X}_B \vert \mathbf{X}_C$
\\for any partition $A,B,C$ of $\left\{1,\ldots,p\right\}$.
\end{assb}
$\mathbf{X}$ having full support with respect to the product of the domains of the individual $X_j$ is sufficient for the intersection property and hence uniqueness of the Markov boundary. Strictly weaker, necessary and sufficient conditions are discussed by \cite{peters2015intersection}.

One estimation strategy would be to consider the individual hypothesis and output the collection of all variables for which these individual hypotheses are true
\begin{align}\label{eq:H0j-ind}
\begin{split}
H_{0,j}: \quad \mathcal{E} & \perp X_j \vert \mathbf{X}_{-j}, \quad \text{where} \quad \mathcal{E} = Y - \EE\left[Y\vert \mathbf{X} \right]\\
\tilde{W} &= \left\{j: H_{0,j} \ \text{is true}\right\}.
\end{split}
\end{align}
We can relate this to the Markov boundary.

\begin{theo}\label{theo:WW}
Assume the model \eqref{eq:SEM}. Let $W$ be any set as in \eqref{eq:mb} and $\tilde{W}$ as in \eqref{eq:H0j-ind}. Then, $W \subseteq \tilde{W}$. If the intersection property \ref{ass:inter} holds, $W=\tilde{W}$, and $W$ is unique.
\end{theo}

In general, \ref{ass:markov} and \ref{ass:add-sep} do not imply that the ANM \eqref{eq:ANM} with only $\mathbf{X}_U$ as predictors is causally well-specified. Therefore, this set cannot be found by looping over all subsets of $\mathbf{X}$ and testing \eqref{eq:H0}.

Recall the examples in Figure \ref{fig:DAGs}. In the left structure, $W=\left\{2\right\}$ if \ref{ass:add-sep} holds for $X_2$. But, $X_2 \rightarrow Y$ is not a causally well-specified ANM unless faithfulness is violated, i.e. $X_2 \perp H$. Similarly, on the right, it holds $W=\left\{1\right\}$ if \ref{ass:add-sep} holds for $X_1$. But, $X_1 \rightarrow Y$ is not a causally well-specified ANM unless faithfulness is violated, i.e. $X_1 \perp H$.

Note also that there is an unobserved causal path \citep{maeda2021causal} from $X_1$ to $Y$. This means that their method which exploits different potential parental sets to obtain independent residuals cannot identify this edge. Nevertheless, the edge can be characterized as causally well-specified when considering conditional independence criteria.

We emphasize that the characterizations in this Section \ref{sec:causal-wellspec} provide the fundamental basis to define the concepts of global and local causal well-specification. This then enables the construction of algorithms that aim to estimate causal well-specification based on finite sample observational data, as discussed next.

\section{Estimating the set of well-specified predictor variables}\label{estim}
We subsequently focus on a specific method to assess conditional dependence. Of course, different estimators could be used as well. The intuition of how conditional independence relates to causal well-specification stays the same. The practical algorithm to estimate the set of variables with well-specified effect is given in Section \ref{multi-alg}.

Throughout this section, we assume that we have n i.i.d.\ observations $\mathbf{x}_1, \ldots, \mathbf{x}_n$ and $y_1, \ldots, y_n$ of $\mathbf{X}$ and $Y$ respectively. More compactly, this data can be written as $\mathbf{x} = \begin{pmatrix}
\mathbf{x}_1, \ldots, \mathbf{x}_n
\end{pmatrix}^\top\in \mathbb{R}^{n \times p}$ and $\mathbf{y} =\begin{pmatrix}
y_1, \ldots, y_n
\end{pmatrix}^\top \in \mathbb{R}^n$. Also, define the unobserved $\epsilon_i = y_i - \EE\left[Y \vert \mathbf{X}=\mathbf{x}_i\right]$.

\subsection{Making use of FOCI (Feature Ordering by Conditional Independence)}
One estimation strategy would be to test the hypotheses in \eqref{eq:H0j-ind} for all $j$. Conditional independence testing is a hard problem on its own; see, e.g., \cite{shah2020hardness}. Here, it is even more challenging as we need to rely on estimated residuals rather than the error terms directly. Instead of testing, we use FOCI (Feature Ordering by Conditional Independence) by \cite{azadkia2021simple}. This method estimates a Markov blanket, which they call a sufficient set, of a target variable, and they give guarantees which hold with high probability for large enough sample size. Thus, it can find a superset of the Markov boundary of $\mathcal{E}$, say, $\hat{S}$, such that $\left(\left\{1, \ldots, p\right\} \setminus \hat{S}\right) \subseteq W$. 

Before reviewing the most important concepts of FOCI, we emphasize what our contribution to the subsequent results is. Here, we need to deal with the harder problem of applying FOCI to the estimated residuals $\hat{\boldsymbol{\epsilon}}$ instead of the true, unobserved residuals $\boldsymbol{\epsilon}$. We extend the theory from \cite{azadkia2021FOCI} to this case and provide asymptotic guarantees in Section \ref{asym}. As additional assumptions, we require only a weak form of consistency for the regression estimates as well as continuous residuals. An example demonstrating the pitfalls of discrete residuals is given in Section \ref{disc}. Furthermore, we show a new result for transforming the data before applying FOCI; see Proposition \ref{prop:discrete}. Finally, we suggest an algorithm that yields more stable estimates in Section \ref{multi-alg}.

We provide now some background on FOCI by focusing on its main concepts. 
The precise definitions can be found in Appendix \ref{app:FOCI}. Assume we want to consider if
\begin{equation*}
\mathcal{E} \perp \mathbf{X}_U \vert \mathbf{X}_S.
\end{equation*}
\cite{azadkia2021FOCI}
define a coefficient of conditional dependence, $T\left(\mathcal{E}, \mathbf{X}_U \vert \mathbf{X}_S\right) \in \left[0,1\right]$, which is $0$ for conditional dependence, $1$ if $\mathcal{E}$ is almost surely a function of $\mathbf{X}_U$ given $\mathbf{X}_S$, and in between otherwise. A slightly different coefficient with the same properties is defined for empty conditioning sets. $T\left(\cdot\right)$ can be decomposed as
\begin{equation*}
T\left(\mathcal{E}, \mathbf{X}_U \vert \mathbf{X}_S\right) = \dfrac{Q\left(\mathcal{E}, \mathbf{X}_U \vert \mathbf{X}_S\right)}{S\left(\mathcal{E}, \mathbf{X}_S\right)},
\end{equation*}
with a non-negative numerator and denominator. Thus, conditional independence is also equivalent to $Q\left(\mathcal{E}, \mathbf{X}_U \vert \mathbf{X}_S\right) = 0$.  By construction
\begin{align*}
Q\left(\mathcal{E}, \mathbf{X}_U \vert \mathbf{X}_S\right) = Q\left(\mathcal{E}, \left\{\mathbf{X}_U,  \mathbf{X}_S\right\}\right) - Q\left(\mathcal{E}, \mathbf{X}_S\right)
\end{align*}
using the version of $Q\left(\cdot\right)$ without a conditioning set. FOCI is designed to greedily find an additional covariate that maximizes $T\left(\mathcal{E}, X_j \vert \mathbf{X}_S\right)$ assuming $\mathbf{X}_S$ has already been chosen. This is equivalent to greedily maximizing $Q\left(\mathcal{E}, \left\{X_j,  \mathbf{X}_S\right\}\right)$ as the normalization $S\left(\cdot\right)$ above does not depend on the candidate variable. For practical evaluation, one can use a sample estimate $Q_n\left(\boldsymbol{\epsilon}, \mathbf{x}_S\right)$. This only depends on the relative order of $\boldsymbol{\epsilon}$ and is not guaranteed to be non-decreasing when adding additional covariates to $S$. Hence, FOCI stops when no candidate variable yields and improvement in $Q_n\left(\cdot\right)$, i.e.,
\begin{equation*}
\forall j \in \left\{1,\ldots,p\right\}\setminus S: \quad Q_n\left(\boldsymbol{\epsilon}, \left\{\mathbf{x}_j,\mathbf{x}_S\right\}\right) \leq Q_n\left(\boldsymbol{\epsilon}, \mathbf{x}_S\right).
\end{equation*}
For large enough data, $Q_n\left(\boldsymbol{\epsilon}, \left\{\mathbf{x}_j,\mathbf{x}_S\right\}\right) \approx Q\left(\mathcal{E}, \left\{X_j,  \mathbf{X}_S\right\}\right)$ such that the algorithm does not stop before the estimated set $\hat{S}$ is a Markov blanket, i.e., it includes all necessary covariates $\left\{1, \ldots, p\right\} \setminus W$ with high probability. However, there is in general no guarantee against superfluous inclusion to $\hat{S}$ and we get $\left(\left\{1, \ldots, p\right\} \setminus \hat{S}\right) \subseteq W$.

For power purposes, it can be advantageous
to consider a certain non-monotonic transformation $g\left(\mathcal{E}\right)$ as input to FOCI. Intuitively, $T\left(g\left(\mathcal{E}\right), X_j \vert \mathbf{X}_S\right)$ measures nonparametrically how much $X_j$ increases the explicative power for $g\left(\mathcal{E}\right)$. Hence, transforming $\mathcal{E}$ such that this relative explicative power increases, makes detection easier. In particular, we suggest the absolute value function. For this, we provide a precise result for symmetric data below. Although exact symmetry is hardly the case except for toy examples, the intuition is that the dependence of $\mathcal{E}$ on $\mathbf{X}$ can be mainly in the second moment, i.e., the scale. Hence, the absolute value transform is then beneficial. For our general results, we assume that $g\left(\cdot\right)$ is an $l$-Lipschitz function whose level sets have Lebesgue measure $0$.

For the precise definitions for $T\left(\cdot\right)$ and $Q\left(\cdot\right)$; see (2.1) and (11.1) in \citep{azadkia2021simple} or Appendix \ref{app:FOCI} here. FOCI greedily increases the set of predictors to maximize $Q$.
\begin{prop}\label{prop:abs}
Let $S \subseteq \left\{1, \ldots, p \right\}$. If $\mathcal{E}\vert \mathbf{X}_S$ has a continuous and symmetric (around 0) distribution, it holds
\begin{equation*}
T\left(\left\vert \mathcal{E}\right\vert, \mathbf{X}_S\right) = 4 T\left(\mathcal{E}, \mathbf{X}_S\right)\quad \text{and} \quad Q\left(\left\vert \mathcal{E}\right\vert, \mathbf{X}_S\right) = 4 Q\left(\mathcal{E}, \mathbf{X}_S\right).
\end{equation*}
\end{prop}
These larger population values can improve the algorithm's performance.

In general, we require some sort of consistency for our regression estimates and our discussion allows any reasonable choice of regression (machine learning) techniques. While as in the population case rejections of the null hypothesis could only be due to hidden confounding or additively non-separable functions, one must always consider insufficient explicative power of the applied regression (machine learning) method as a further reason in the finite sample case.

We consider two different algorithms.

\begin{algorithm}[h!]
\caption{In-sample FOCI}\label{alg:in}
\hspace*{\algorithmicindent} \textbf{Input} i.i.d.\ data $\mathbf{x} \in \mathbb{R}^{n \times p}$ and $\mathbf{y} \in \mathbb{R}^n$, and function $g\left(\cdot\right)$\\
\hspace*{\algorithmicindent} \textbf{Output} estimated set of variables $\hat{S}$ for which null hypothesis \eqref{eq:H0j} is rejected 
\begin{algorithmic}[1]
\State Get an estimate $\hat{f}\left(\mathbf{X}\right)$ for $\EE\left[Y\vert \mathbf{X}\right]$ using a certain regressor
\State Estimate the residuals as $\hat{\boldsymbol{\epsilon}} = \mathbf{y} - \hat{f}\left(\mathbf{x}\right)$
\State  Apply FOCI \citep{azadkia2021simple} to the data $\left(g\left(\hat{\boldsymbol{\epsilon}}\right), \mathbf{x}\right)$ to get the set $\hat{S}$
\end{algorithmic}
\end{algorithm}

\begin{algorithm}[h!]
\caption{Sample splitting FOCI}\label{alg:split}
\hspace*{\algorithmicindent} \textbf{Input} i.i.d.\ data $\mathbf{x} \in \mathbb{R}^{n \times p}$ and $\mathbf{y} \in \mathbb{R}^n$, and function $g\left(\cdot\right)$\\
\hspace*{\algorithmicindent} \textbf{Output} estimated set of variables $\hat{S}$ for which null hypothesis \eqref{eq:H0j} is rejected 

\begin{algorithmic}[1]
\State Split the data uniformly at random into two disjoint parts of sizes $\lfloor n/2 \rfloor$ and $\lceil n/2 \rceil$, say, $\left(\mathbf{x}^{\left(1\right)}, \mathbf{y}^{\left(1\right)}\right)$ and $\left(\mathbf{x}^{\left(2\right)}, \mathbf{y}^{\left(2\right)}\right)$
\State Get an estimate $\hat{f}\left(\mathbf{X}\right)$ for $\EE\left[Y\vert \mathbf{X}\right]$ using a certain regressor on the data $\left(\mathbf{x}^{\left(1\right)}, \mathbf{y}^{\left(1\right)}\right)$
\State Estimate the residuals as $\hat{\boldsymbol{\epsilon}}^{\left(2\right)} = \mathbf{y}^{\left(2\right)} - \hat{f}\left(\mathbf{x}^{\left(2\right)}\right)$
\State  Apply FOCI \citep{azadkia2021simple} to the data $\left(g\left(\hat{\boldsymbol{\epsilon}}^{\left(2\right)}\right), \mathbf{x}^{\left(2\right)}\right)$ to get the set $\hat{S}$
\end{algorithmic}
\end{algorithm}

For notational simplicity, we call the data that is input to FOCI $\left(\hat{\boldsymbol{\epsilon}}, \mathbf{x}\right)$ in our theoretical derivations regardless of the applied algorithm, i.e., we omit the superscript in the splitting case. The advantage of Algorithm \ref{alg:split} is that the residuals estimated on the hold-out split are still i.i.d.\ which simplifies things, at least analytically. Furthermore, the sample splitting idea enables further favourable algorithms to be presented in Section \ref{multi-alg}.

\subsection{Asymptotic results}\label{asym}
We generally make the following assumptions for applying FOCI to an estimated $\hat{\boldsymbol{\epsilon}}$.

\begin{assb}\label{ass:suitable}
$\left\vert \hat{\epsilon}_i - \epsilon_i \right\vert= {\scriptstyle \mathcal{O}_p}\left(1\right)$.
\end{assb}
\begin{assb}\label{ass:continuous}
$\mathcal{E}$ is a continuous random variable.
\end{assb}
\begin{assb}\label{ass:dep}
$\nexists S \subseteq \left\{1, \ldots, p \right\}$ such that $\mathbf{X}_{-S} \not \perp \mathcal{E} \vert \mathbf{X}_S$ but $\mathbf{X}_{-S} \perp g\left(\mathcal{E}\right) \vert \mathbf{X}_S$.
\end{assb}
The probability in \ref{ass:suitable} is with respect to both the regression estimate and the new data point $i$. The assumption is slightly different depending on which algorithm is applied. Apart from invoking \ref{ass:continuous} for the proofs, we provide a simple example in Section \ref{disc} to show that discrete distributions can lead to inconsistency.

The main proof ingredient for adapting the results to our setting is showing that for random indices $i$ and $l$ the probability that the estimated residuals imply a different relative ordering than the true residuals approaches $0$. 

With sample splitting, we obtain a consistency result analogous to \cite{azadkia2021simple}. We require the same regularity conditions as they do.
Let
\begin{equation*}
\delta = \underset{j,S: \ T\left(g\left(\mathcal{E}\right), X_j \vert \mathbf{X}_S \right) > 0}{\min} Q\left(g\left(\mathcal{E}\right), \left\{\mathbf{X_S}, X_j\right\}\right) - Q\left(g\left(\mathcal{E}\right), \mathbf{X_S}\right),
\end{equation*}
i.e., the lowest difference in $Q\left(\cdot\right)$ we should be able to detect.
\begin{assapp}\label{ass:az1}
There are nonnegative real numbers $\beta$ and $C$ such that for any set $S\subseteq \{1,\ldots,p\}$ of size $s \leq 1/\delta+2$, any $\mathbf{x}_S,\mathbf{x}_S'\in \mathbb{R}^s$ and any  $t\in \mathbb{R}$,
\begin{align*}
&\left\vert P(g\left(\mathcal{E}\right)\ge t \vert \mathbf{X} = \mathbf{x}_S)  - P(g\left(\mathcal{E}\right)\ge t\vert \mathbf{X} = \mathbf{x}'_S)\right\vert\\
&\le C(1+\Vert\mathbf{x}_S\Vert^\beta+\Vert\mathbf{x}'_S\Vert^\beta) \Vert\mathbf{x}_S-\mathbf{x}'_S\Vert.
\end{align*}
\end{assapp}

\begin{assapp}\label{ass:az2}
There are positive numbers $C_1$ and $C_2$ such that for any $S$ of size $s \le 1/\delta+2$ and any $t>0$, $\PP(\left\Vert\mathbf{X}_S\right\Vert\ge t)\le C_1 e^{-C_2t}$. 
\end{assapp}

\begin{theo}\label{theo:cons}
Suppose that the regularity assumptions \ref{ass:az1} and \ref{ass:az2} \citep{azadkia2021simple} for the data $\left(g\left(\mathcal{E}\right), \mathbf{X}\right)$ hold as well as conditions \ref{ass:inter} - \ref{ass:dep}. Let $\hat{S}$ be the output of Algorithm \ref{alg:split}. There are positive real numbers $L_1$, $L_2$ and $L_3$ that do not depend on the sample size such that
\begin{equation*}
\PP\left(\hat{S} \supseteq \left\{1,\ldots,p\right\} \setminus W\right) \geq 1 - L_1 p^{L_2} \exp\left(-L_3 n\right).
\end{equation*}
\end{theo}

Without sample splitting, $\left(g\left(\hat{\boldsymbol{\epsilon}}\right),\mathbf{x}_U\right)$ are not independent copies. Therefore, the bounded difference inequality \citep{mcdiarmid1989method} which is applied to obtain the exponential probability decay cannot be used. Nevertheless, convergence in probability is still true.
\begin{theo}\label{theo:cons2}
Assume the conditions of Theorem \ref{theo:cons}. Let $\hat{S}$
be the output of Algorithm \ref{alg:in}. Then,
\begin{equation*}
\underset{n \rightarrow \infty}{\lim}\PP\left(\hat{S} \supseteq \left\{1,\ldots,p\right\} \setminus W\right) = 1.
\end{equation*}
\end{theo}
This result is derived by a simple application of the Markov inequality instead of the bounded difference inequality. As the $\left(g\left(\hat{\boldsymbol{\epsilon}}\right),\mathbf{x}_U\right)$ become decreasingly dependent from another with increasing sample size, we conjecture that the true convergence rate could be similar to the one for sample splitting.

\subsubsection{Discrete $\mathcal{E}$}\label{disc}
For our results, we invoked assumption \ref{ass:continuous}, i.e., the residuum is a continuous random variable. A simple toy example shows that a discrete random variable might invalidate the asymptotic guarantees. Use the definition (2.1) in \cite{azadkia2021simple} for $T\left(\cdot\right)$, i.e., $T\left(\mathcal{E},X\right)=0$ if and only if  $\mathcal{E}$ and $X$ are independent. Let $T_n\left(\cdot\right)$ be its suggested sample estimate; see also Appendix \ref{app:FOCI}.
\begin{prop}\label{prop:discrete}
Let $X$ be a bounded, continuous random variable and $\mathcal{E}$ a centered random variable that is uniformly distributed over a discrete set of size $k>1$ independent from $X$ such that $T\left(\mathcal{E}, X\right) = 0$. Let $Y \leftarrow X \beta + \mathcal{E}$ for some $\beta \neq 0$. Apply linear least squares regression, which fulfils \ref{ass:suitable}, to $n$ i.i.d.\ copies $\left(\mathbf{y}, \mathbf{x}\right)$ to get the estimates $\hat{\boldsymbol{\epsilon}}$. It holds
\begin{equation*}
\EE\left[T_n\left(\hat{\boldsymbol{\epsilon}}, \mathbf{x}\right)\right] \overset{n \rightarrow \infty}{\rightarrow} \dfrac{1}{k^2}>0.
\end{equation*}
\end{prop}

We provide some intuition while the detailed proof can be found in Appendix \ref{app:proof-discrete}. As $\mathbf{x}$ is continuous, it is never perfectly orthogonal to $\boldsymbol{\epsilon}$, and the least squares estimator is slightly off. Then, within each of the $k$ groups the ranking of the $\hat{\boldsymbol{\epsilon}} = \boldsymbol{\epsilon} + \mathbf{x} \left(\beta - \hat{\beta}\right)$ is exactly according to the ranking of the $\mathbf{x}$ or inverted such that there is some non-vanishing dependence that FOCI detects.

\subsection{Practical algorithm}\label{multi-alg}
Although we can consistently find a Markov blanket (but not necessarily the minimal Markov boundary) using Algorithms \ref{alg:in} or \ref{alg:split} as the sample size grows, there are several drawbacks to that. First, there is no protection against including superfluous variables into $\hat{S}$ and typically this happens with non-negligible probability. Second, for low sample sizes, $\hat{S}$ can miss out on some variables.

To partially remedy these issues, we incorporate ideas from multisplitting \citep{meinshausen2009p} and stability selection \citep{meinshausen2010stability}. We apply Algorithm \ref{alg:split} repeatedly with several random data partitions. Inspired by \cite{shah2013variable} who suggest using ``complementary pairs'', i.e., both halves of every split, we let each halve be used once for estimating the conditional mean and once for independence testing.

\begin{algorithm}[b!]
\caption{Selection of variables with well-specified effect using multiple splits}\label{alg:multisplit+test}
\hspace*{\algorithmicindent} \textbf{Input} i.i.d.\ data $\mathbf{x} \in \mathbb{R}^{n \times p}$ and $\mathbf{y} \in \mathbb{R}^n$, function $g\left(\cdot\right)$, number of repetitions $B$, and significance levels $\alpha$ and $\tilde{\alpha}$. \\
\hspace*{\algorithmicindent} \textbf{Output} estimated set of variables $\hat{W}$ with causally well-specified effect 
\begin{algorithmic}[1]
\State $n_j = 0 \ \forall j=1,\ldots,p$
\For{$b=1$ to $B$}
\State Split the data uniformly at random into two disjoint parts of sizes $\lfloor n/2 \rfloor$ and $\lceil n/2 \rceil$, say, $\left(\mathbf{x}^{\left(1\right)}, \mathbf{y}^{\left(1\right)}\right)$ and $\left(\mathbf{x}^{\left(2\right)}, \mathbf{y}^{\left(2\right)}\right)$
\State Get an estimate $\hat{f}\left(\mathbf{X}\right)$ for $\EE\left[Y\vert \mathbf{X}\right]$ using a certain regressor on the data $\left(\mathbf{x}^{\left(1\right)}, \mathbf{y}^{\left(1\right)}\right)$
\State Estimate the residuals as $\hat{\boldsymbol{\epsilon}}^{\left(2\right)} = \mathbf{y}^{\left(2\right)} - \hat{f}\left(\mathbf{x}^{\left(2\right)}\right)$
\State Apply the HSIC test to the data $\left(\hat{\boldsymbol{\epsilon}}^{\left(2\right)}, \mathbf{x}^{\left(2\right)}\right)$ to get the p-value $p^b$.
\State Apply FOCI \citep{azadkia2021simple} to the data $\left(g\left(\hat{\boldsymbol{\epsilon}}^{\left(2\right)}\right), \mathbf{x}^{\left(2\right)}\right)$ to get the set $\hat{S}^b$
\State Swap the roles of $\left(\mathbf{x}^{\left(1\right)}, \mathbf{y}^{\left(1\right)}\right)$ and $\left(\mathbf{x}^{\left(2\right)}, \mathbf{y}^{\left(2\right)}\right)$ and repeat the previous steps to get $p^{B + b}$ and $\hat{S}^{B+b}$
\EndFor
\State Combine the p-values $p^1, \ldots p^{2B}$ \citep{meinshausen2009p} and get the model p-value $p^0$.
\If{$p^0 > \alpha$}
\State $\hat{W} = \left\{1,\ldots,p\right\}$
\Else
\State $\hat{W} = \emptyset$
\For{$j=1$ to $p$}
\State $n_j = \sum_{b = 1}^{2B} j \in \hat{S}^b$
\EndFor
\State $\bar{n} = \sum_{j=1}^{p} n_j / p$
\State $n^{\min}=\underset{j: n_j \geq \bar{n}}{\min n_j}$
\For{$j=1$ to $p$}
\If{$n_j < \bar{n}$ and $proportion.test\left(n_j, n^{min}, 2B\right) \leq \tilde{\alpha}$}
\State $\hat{W} = \hat{W} \cup j$
\EndIf
\EndFor
\EndIf
\end{algorithmic}
\end{algorithm}

As unconditional independence is easier to assess than conditional independence, we first test for $H_0$ as in \eqref{eq:H0}. For this, we apply the test by \cite{pfister2018kernel}. The case where only estimates of the residuals are available is explicitly discussed in their work. Then, we combine the p-values over the different splits as suggested by \cite{meinshausen2009p}. Only if the global model is rejected, the individual covariates are inspected.

If we cannot trust the overall model, we only consider the effects of variables that are selected substantially less than others by the FOCI algorithm to be causally well-specified. We split the variables into two groups: those that are selected by FOCI below average over the splits and the others. For the latter group, we reject $H_{0,j}$. Each variable from the first group we compare to the least selected variable from the second group with some proportion test such as Fisher's exact test. The variables that show significant differences are added to the estimated well-specified set $\hat{W}$. Notably, there is no exact interpretation of the significance level used for these tests, but the intuition that a lower significance level leads to fewer false positives in the set $\hat{W}$ remains true. In contrast, a lower significance level for the preceding test of the global model leads to the methods becoming more liberal.

The intuition behind splitting at the mean is the following.
For large enough sample size, the necessary variables are selected by FOCI in almost every split, see Theorem \ref{theo:cons}, while the variables with causally well-specified effect could be selected with some probability much lower than $1$. The mean separates the two groups and there is a significant difference in the selection fraction of the two groups. For a low sample size, the behaviour of FOCI is more random. However, as long as no variables stand out, we do not add any to $\hat{W}$, i.e., if $H_{0,j}$ is not true, the probability $\PP\left(j \in \hat{W}\right)$ is moderately low. However, it is lower bound by the type II error of the global test. This is fundamental to our idea. If the sample size is such that the global test, i.e., unconditional independence testing, does not work well yet, the local analysis is also not of much use.

We summarize the procedure in Algorithm \ref{alg:multisplit+test}.

\section{Simulation example}\label{sim}
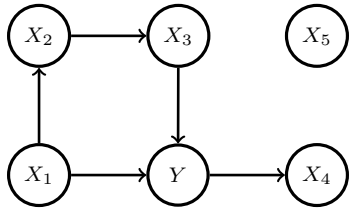
\begin{figure}[h!]
\centering
\begin{tikzpicture}[roundnode/.style={circle, very thick, minimum size=8mm}]
\node[draw, roundnode, text centered] (x1) {${\scriptstyle X_1}$};
\node[draw, roundnode, above =of x1, text centered] (x2) {${\scriptstyle X_2}$};
\node[draw, roundnode, right =of x2, text centered] (x3) {${\scriptstyle X_3}$};
\node[draw, roundnode, right =of x1, text centered] (y) {${\scriptstyle Y}$};
\node[draw, roundnode, right =of y, text centered] (x4) {${\scriptstyle X_4}$};
\node[draw, roundnode, above =of x4, text centered] (x5) {${\scriptstyle X_5}$};

\draw[->, line width= 1] (x1) -- (x2);
\draw[->, line width= 1] (x1) -- (y);
\draw[->, line width= 1] (x2) -- (x3);
\draw[->, line width= 1] (x3) -- (y);
\draw[->, line width= 1] (y) -- (x4);
\end{tikzpicture}
 \caption[Sample DAGs]
 {DAG representing the SCM in the simulation.}
 \label{fig:DAG-sim}
\end{figure}

We evaluate the method on a simple SCM represented by the DAG in Figure \ref{fig:DAG-sim}. We let the causal effects be non-monotonic functions. As discussed in \cite{schultheiss2023pitfalls}, non-monotonic effects can lead to stronger dependence between residual and predictor in the wrong direction. Hence, we are a bit more sample-efficient than with monotonic functions. The effects have the following form
\begin{equation*}
f\left(X_j\right) = \alpha_1 \left\vert X_j\right \vert ^{\beta_1} \text{sign}\left(X_j\right) + \alpha_2 \left\vert X_j \right \vert ^{\beta_2},
\end{equation*}
where the parameters are randomly sampled and differ for every simulation run. The causal effect on $Y$ is additive in the parents. We standardize and normalize the effects. The additive error terms are either normal, uniform, or Laplace with variance $1$ for the root nodes and $1/4$ for the others. The different distributions are randomly assigned to the different nodes; two of each.

We consider all possible subsets of size $3$ as observed predictors. Denote this observed subset by $M$. For $M=\left\{1,2,3\right\}$ and $M=\left\{1,3,5\right\}$ the additive noise model is causally well-specified.

We consider $100$ different random setups for sample sizes $10^2$ to $10^5$. For each, we consider all possible $M$. To get $\hat{W}$ we apply Algorithm \ref{alg:multisplit+test} with $B=25$ splits and the absolute value function as $g\left(\cdot\right)$.

For the regression, we apply eXtreme Gradient Boosting implemented in the \textsf{R}-package \texttt{xgboost} \citep{chen2021xgboost}. Other regression (machine learning) techniques could be used instead if they are flexible enough. We fix our choice here for this proof of concept. We use the respective left-out split of the data for early stopping when fitting the regression functions. This is a slight violation of our theoretical algorithm where the residuals are perfectly independent.  We use the authors' implementation of FOCI \citep{azadkia2021FOCI} and dHSIC \citep{pfister2019dHSIC}.

\begin{figure}[b!]
 \centering
 \includegraphics[width=1\textwidth]{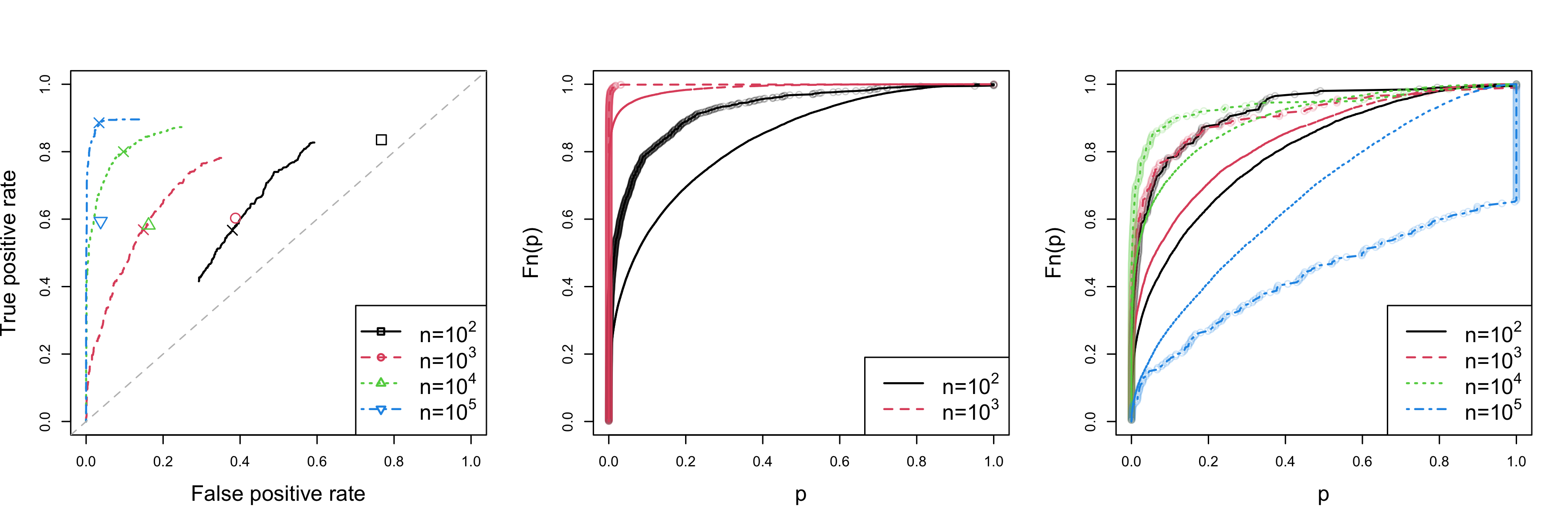}
 \caption[Simulation in a SEM]
 {The results are based on $100$ simulation runs. On the left: False positive rate versus true positive rate obtained with Algorithm \ref{alg:multisplit+test} for varying $\tilde{\alpha}$ and $\alpha = 0.05$. The crosses correspond to $\tilde{\alpha} = 0.01$. The other symbols describe the performance of Algorithm \ref{alg:split}. In the middle: empirical cumulative distribution function of the p-values obtained with HSIC. We compare the raw p-values from each split (lines only) to the cumulated p-value per simulation run (lines with dots). On the right:  the same for the models fulfilling $H_0$.}
 \label{fig:rand-RE}
\end{figure}

For the predictor sets where global causal well-specification does not hold, we consider the false positive rate (FPR) $\hat{\PP}\left(j \in \hat{W}\vert j \not \in W\right)$ and the true positive rate (TPR) $\hat{\PP}\left(j \in \hat{W}\vert j \in W\right)$ for adding predictors to the set $\hat{W}$. Here, $\hat{\PP}\left(\cdot\right)$ denotes the empirical probability over our simulation runs.  We fix $\alpha=0.05$ and consider varying values of $\tilde{\alpha}$. The resulting rates are on the left in Figure \ref{fig:rand-RE}.

For $n=10^2$, a low FPR is not attainable because the p-values for \eqref{eq:H0} are not reliably small enough, and Algorithm \ref{alg:multisplit+test} often terminates before considering the individual covariates. However, even for this low sample size, we get a performance that is clearly better than random guessing. For large sample sizes, the FPR becomes very low which is in agreement with Theorem \ref{theo:cons}. The lack of power is mainly due to the subsets of predictors with $\left\vert W \right\vert > 1$. FOCI chooses superfluous covariates with non-vanishing probability for every sample size. Hence, the two covariates with causally well-specified effects may be selected with a frequency that differs a lot between the two. If one then appears to be more similar to the covariate with not well-specified effect, our algorithm misses out on this such that $\hat{W} \subset W$.

For comparison, we also show the results if we instead only consider a single random split where $50\%$ of the data is used to estimate the residuals and the other $50\%$ to assess independence. If $H_0$ is rejected we apply Algorithm \ref{alg:split} (using the same splits) and choose $\hat{W}$ to be the complement of the set chosen by FOCI. Except for $n=10^2$, this lies below the curve for multiple splits, i.e., there is an $\tilde{\alpha}$ that is better in terms of both FPR and TPR. Further, our default choice $\tilde{\alpha} = 0.01$ is more conservative. For large enough sample size, using $\tilde{\alpha}=0.01$ leads to more power than considering a single split. Hence, even though the problem is hard in general, aggregating information over multiple random splits of the same dataset can lead to a performance boost.

We also evaluate the testing of $H_0$ \eqref{eq:H0}. For this, we show the empirical cumulative distribution function of the obtained p-values in the middle of Figure \ref{fig:rand-RE}. We consider the p-value aggregated over the splits as well as the individual p-values considering single splits. For the largest sample sizes, the distribution of both is visibly not distinguishable from a point mass at $0$. We omit this in the plot for the sake of overview. For $n=10^2$ and $n=10^3$, aggregating the p-values over splits helps to reject the global model for most possible significance levels. The acceptance rate for the global model poses a lower bound to the attainable FPR for every subsequent per-covariate analysis. For $n=10^2$ and $\alpha = 0.05$, this rate is around $0.56$ for single splits and reduced to roughly $0.33$ by aggregating. This confirms the usefulness of the multisplitting idea.

We also consider the distribution of the p-values for the two subsets of predictors that yield causally well-specified models. This is shown on the right side of Figure \ref{fig:rand-RE}. We see that the raw p-values are too liberal. By construction, this effect is enhanced by aggregation over the splits. For increasing sample size, there are two competing effects. The regression approximation becomes better leading to less dependent residuals. But, the tests become more powerful in detecting spurious dependence. As the HSIC implementation cannot handle $5 \cdot 10^4$ samples, we only test with $10^4$ samples per split. Hence, the p-values for $n=10^5$ are likely more liberal theoretically. In summary, we see that testing for \eqref{eq:H0} is already difficult per se. However, one can also see it the other way around: if the regression is unable to render the residuals independent one should not trust the obtained function even if there was a true underlying ANM.

In this example, fitting only additive functions with no interactions between the measured covariates leads to the same conclusion given perfect regression fit and independence tests since the data follow a CAM \citep{buhlmann2014cam}. Hence, if one restricts the analysis to additive functions due to pre-knowledge or just by assumption the problem could become easier. When applying GAM regression as implemented in \texttt{mgcv} \citep{wood2011fast}, the results for the causally not well-specified predictor sets remain qualitatively similar. The p-values for the models fulfilling $H_0$ are still visibly clearly not uniformly distributed. But, they become less liberal. This is as finding the true conditional mean and hence the true independent residuals becomes easier. For $n=10^5$ the distribution of the raw p-values is sufficiently close to uniform such that the aggregated p-values are even super-uniform. Again, this needs to be taken with a grain of salt as not all samples can be used for testing independence.

\section{Real data analysis}
We consider the K562 dataset provided by \cite{replogle2022mapping}. We follow the preprocessing in the benchmark of \cite{chevalley2022causalbench}. Then, the dataset contains 162,751 measurements of the activity of 622 genes: 10,691 of the measurements are taken in a purely observational environment while the remaining are obtained under various interventions. For each gene, there exists an environment where it has been intervened on by a knockdown using CRISPRi \citep{larson2013crispr}, i.e., its activity is reduced. As our method is designed for i.i.d.\ data, we only consider the observational environment henceforth. With the interventions, some sanity checks of our findings are possible as discussed below.

We make a pre-selection of the measured covariates 
before applying our method. There are $28$ genes that are active, i.e., greater than $0$, in each measurement in the observational sample. We restrict our analysis to these and call them $X_1$ to $X_{28}$ for simplicity. Within these $28$, we estimate Markov blankets using FOCI. For each of the $28$ genes, two estimates are implied: all the genes selected by FOCI when this covariate is the target as well as all the genes for which this covariate is in the output of FOCI. As
the target, say $Y$, we choose the one with the highest agreement between the two estimated sets in terms of intersection size relative to the size of the union.  For the target $Y$, we then consider the intersection of the Markov blankets mentioned above (where $Y$ is the target or appears in the output of FOCI).
This results in three predictors, $X_{10}$, $X_{12}$, and $X_{15}$.

With the selected target and predictors we run Algorithm \ref{alg:multisplit+test} with $B = 25$ splits using \texttt{xgboost} for regression. There is a strong indication against the global null hypothesis \eqref{eq:H0} with a p-value of roughly $10^{-27}$. Hence, we proceed to the per-covariate analysis. Covariate $X_{15}$ is in $\hat{S}^b$ 41 out of 50 times while as for the others it is only 22 ($X_{10}$) and 19 ($X_{12}$). Hence, the effects of the latter appear to be causally well-specified and we get the set $\hat{W}=\left\{X_{10}, X_{15}\right\}$ when running Algorithm \ref{alg:multisplit+test} with our suggested default of $\tilde{\alpha}=0.01$.

To assess the success of our method, we now consider the available interventional data.  Comparing the distribution of $X_k$ when the activity of $X_j$ is reduced by an external intervention to its observational distribution, gives an assessment of whether there is a causal effect from $X_j$ to $X_k$. We do this using a Mann-Whitney U test. Intervening on any of the three predictor covariates appears to highly influence the activity of $Y$ with p-values of the order $10^{-4}$, $10^{-13}$, and $10^{-6}$. In the reverse direction, intervening on $Y$ does not have strong influence on $X_{10}$ ($p\approx 0.1$) and $X_{12}$ ($p \approx 0.5$) but on $X_{15}$ ($p \approx 4*10^{-5}$). Thus, there appears to be some cyclic effect between $Y$ and $X_{15}$. Hence, it is less appropriate to consider its regression effect to be causally well-specified whereas our estimated well-specification for $X_{10}, X_{12}$ on $Y$ is compatible with the validation analysis based on interventional data.

\begin{table}[b!]
\centering
\begin{tabular}{|c|c|c|c|c|c|}
\hline
Target $Y$ & Predictor $X_j$& Mann-Whitney U test & Splits & Proportion test & Relative bias \\
\hline
\multirow{4}{*}{$X_{5}$}
& $X_{2}$ & 2.3e-18 & 14 & 1.2e-02 & 1.3e-01 \\
& $X_{3}$ & 4.9e-31 & 26 & -- & 2.1e-02 \\
& $X_{4}$ & 1.2e-69 & 19 & 1.1e-01 & 3e-02 \\
& $X_{12}$ & 3.5e-01 & 28 & -- & 4.3e-02 \\
\hline
\multirow{2}{*}{$X_{6}$}
& $X_{11}$ & 7.7e-01 & 17 & 2.3e-05 & 1e-01 \\
& $X_{24}$ & 2.4e-09 & 38 & -- & 1.5e-01 \\
\hline
\multirow{4}{*}{$X_{7}$}
& $X_{8}$ & 3.3e-02 & 28 & -- & 1e-01 \\
& $X_{9}$ & 1.4e-16 & 10 & 2e-04 & 8.3e-02 \\
& $X_{14}$ & 1.2e-79 & 30 & -- & 9.7e-02 \\
& $X_{22}$ & 1.6e-35 & 11 & 4.6e-04 & 4.1e-02 \\
\hline
\multirow{3}{*}{$X_{9}$}
& $X_{7}$ & 5.4e-01 & 14 & 2.2e-03 & 1.8e-02 \\
& $X_{11}$ & 1.2e-14 & 30 & -- & 1.8e-02 \\
& $X_{22}$ & 2.3e-06 & 29 & -- & 2.4e-02 \\
\hline
\multirow{2}{*}{$X_{15}$}
& $X_{11}$ & 2.3e-02 & 12 & 4.9e-07 & 1.1e-01 \\
& $X_{16}$ & 1.6e-06 & 37 & -- & 1.2e-01 \\
\hline
\multirow{3}{*}{$X_{16}$}
& $X_{10}$ & 1e-01 & 22 & 7.7e-05 & 4.7e-02 \\
& $X_{12}$ & 4.7e-01 & 19 & 6.3e-06 & 7.7e-02 \\
& $X_{15}$ & 3.8e-05 & 41 & -- & 1.2e-01 \\
\hline
\end{tabular}
\caption{Application to the K562 dataset with varying targets and predictor sets. The third column is the p-value of the Mann-Whitney U test comparing the observational distribution of the predictor to its distribution when knocking down the target. The fourth and fifth column report the output of Algorithm \ref{alg:multisplit+test}, i.e., the number of splits where FOCI selects this predictor, $n_j$, and the p-value of the proportion test if $n_j < \bar{n}$ (the significant findings with small p-value correspond to the variables which are causally well-specified; no p-value indicates that $n_j \geq \bar{n}$ and the variable is not causally well-specified). The last column reports the relative bias $RB^{X_j \rightarrow Y}$ \eqref{eq:RB} when using the model fit on observational data to predict the target in the dataset where the predictor is knocked down.}
\label{tab:data}
\end{table}

Finally, we can also compare how well our regression model trained on the observational data performs on data from the different interventional environments. We do this comparison in terms of absolute bias relative to $Y$'s mean activity in the observational sample, i.e.,
\begin{equation}\label{eq:RB}
RB^{X_j \rightarrow Y}= \dfrac{\left\vert\sum_{i \in \mathcal{D}_j}y_i - \hat{f}\left(\mathbf{x}_i\right)\right\vert / \left\vert \mathcal{D}_j \right\vert}{\sum_{i \in \mathcal{D}_\mathcal{O}}y_i / \left\vert \mathcal{D}_\mathcal{O} \right\vert},
\end{equation}
where $\mathcal{D}_j$ denotes the data points where $X_j$ is knocked down, $\mathcal{D}_\mathcal{O}$ the observational data, and $\hat{f}\left(\cdot\right)$ is trained on $\mathcal{D}_\mathcal{O}$.
This suggests that generalization to the environment where a knockdown is applied to $X_{15}$ works the least with a relative bias \eqref{eq:RB} of about $12\%$ while in the other environments it is roughly $5\%$ or $8\%$ respectively. It must be noted that most data points in the knocked down environments are outside the support of the observational training data such that $\hat{f}\left(\mathbf{X}\right)$ can also be a poor approximation for causal effects; see also the discussion regarding out-of-support interventions in Section \ref{local}. Hence, this analysis of the regression performance in other environments, although in line with our other results, shall be viewed with some caution. The analysis for this target variable corresponds to the last row-box in Table \ref{tab:data}.

Of course, other genes could be viewed as target $Y$. When estimating a Markov blanket as described above for different variables, the interventional environments often indicate the existence of cyclicity between the target and all its potential causes. Then, our method is of little help as the different predictors cannot be grouped into different classes. In Table \ref{tab:data}, we summarize the results for all possible targets with multiple predictors where at least one predictor appears to be neither a descendant of the target nor in a cyclic relation using a threshold of $0.01$ for the Mann-Whitney U test. In $4$ out of $6$ cases, the ranking implied by our method in terms of number of splits where a predictor is selected by FOCI is in agreement with the ranking implied by the Mann-Whitney U test, and $\hat{W}$ using $\tilde{\alpha}=0.01$ is exactly as implied by the interventional data. Of the remaining two cases, the method is once conservative $\hat{W} = \emptyset$ (for $Y = X_5$) and once the interventional data suggest that there are false positives in $\hat{W}$ (for $Y=X_7$). $Y=X_{16}$ is the case discussed in more detail above.

\section{Location-scale noise models}
A simple extension of model \eqref{eq:ANM}, that has recently gained some attention,  is the heteroskedastic noise model also referred to as the location-scale noise model (LSNM). There, the independent, additive noise is scaled by some nonnegative function $g_{\mathbf{X}Y}\left(\mathbf{X}_{\text{PA}\left(Y\right)}\right)$ inducing heteroskedasticity. This is the leading causal model in, e.g., \cite{xu2022inferring, strobl2022identifying, immer2022loci}, where the latter two provide identifiability guarantees.
In analogy to \eqref{eq:ANM}, we call the LSNM causally well-specified if
\begin{equation}\label{eq:het-conditions}
Y \leftarrow f_{\mathbf{X}Y}\left(\mathbf{X}_{\text{PA}\left(Y\right)}\right) + g_{\mathbf{X}Y}\left(\mathbf{X}_{\text{PA}\left(Y\right)}\right) f_{\mathbf{H}Y}\left(\mathbf{H}_{\text{PA}\left(Y\right)}\right), \quad \text{where} \quad \mathbf{H}_{\text{PA}\left(Y\right)} \perp \mathbf{X}.
\end{equation}
We choose the parametrization $\EE\left[f_{\mathbf{H}Y}\left(\mathbf{H}_{\text{PA}\left(Y\right)}\right)\right]=0$ and $\EE\left[f_{\mathbf{H}Y}\left(\mathbf{H}_{\text{PA}\left(Y\right)}\right)^2\right]=1$ such that $f_{\mathbf{X}Y}\left(\cdot\right)$ and $g^2_{\mathbf{X}Y}\left(\cdot\right)$ denote the conditional mean and variance.
As before, the independence condition implies
\begin{equation*}
Y\vert \mathbf{X}=\mathbf{x} \quad \overset{d}{=} \quad Y\vert \text{do}\left(\mathbf{X} \leftarrow \mathbf{x}\right) 
\end{equation*}
With the others, one can separate the independent noise term such that one can understand the counterfactual of changing the predictors.
\begin{align*}
P_Y^{\mathfrak{C} \vert \mathbf{Z}=\mathbf{z}; \text{do}\left(\mathbf{X}\leftarrow \mathbf{x}'\right)}&=\delta_{y'} \quad \text{where}\\
y' &= \left(y -\EE\left[Y\vert \mathbf{X} = \mathbf{x}\right]\right)\sqrt{\text{Var}\left(Y \vert \mathbf{X}=\mathbf{x}'\right)/\text{Var}\left(Y \vert \mathbf{X}=\mathbf{x}\right)} + \EE\left[Y\vert \mathbf{X} = \mathbf{x}'\right].
\end{align*}
To check \eqref{eq:het-conditions}, we have the natural proxy
\begin{equation}\label{eq:H0-het}
H_0: \quad  \mathcal{E} \perp \mathbf{X}, \quad \text{where} \quad \mathcal{E} = \dfrac{Y - \EE\left[Y\vert \mathbf{X} \right]}{\sqrt{\text{Var}\left(Y \vert \mathbf{X}\right)}}
\end{equation}
since under \eqref{eq:het-conditions} we have that $\mathcal{E} = f_{\mathbf{H}Y}\left(\mathbf{H}_{\text{PA}\left(Y\right)}\right)$ and hence $H_0$ in \eqref{eq:H0-het} holds.

In case of model misspecification, we can consider the per-covariate causal well-specification. Condition \ref{ass:markov} remains the same for the LSNM, \ref{ass:add-sep} can be replaced by a weaker version for this more flexible causal model:
\setcounter{assap}{1}
\begin{assap}\label{ass:add-mult-sep}
$Y \leftarrow f_{\mathbf{X}_U Y}\left(\mathbf{X}_U, \mathbf{X}_{\text{PA}\left(Y\right) \setminus U}\right) + g_{\mathbf{X}_U Y}\left(\mathbf{X}_U, \mathbf{X}_{\text{PA}\left(Y\right) \setminus U}\right) f_{\mathbf{H}Y}\left(\mathbf{H}_{\text{PA}\left(Y\right)}, \mathbf{X}_{\text{PA}\left(Y\right) \setminus U}\right)$, i.e., with addition and multiplication of measured functions, one can separate a term that does not include $\mathbf{X}_U$.
\end{assap}
Again, these assumptions imply a counterfactual statement and a testable proxy.

\begin{theo}\label{theo:H0j-het}
Assume the model \eqref{eq:SEM} with \ref{ass:indep}. Let $\mathbf{X}_U$ be a set of covariates fulfilling \ref{ass:markov} and \ref{ass:add-mult-sep}, then
\begin{align*}
&P_Y^{\mathfrak{C} \vert \mathbf{Z}=\mathbf{z}; \text{do}\left(\mathbf{X}_U \leftarrow \mathbf{x}'_U, \mathbf{X}_{-U} \leftarrow \mathbf{X}_{-U}\right)}=\delta_{y'} \quad \text{where}\\
y' = &\left(y -\EE\left[Y\vert \mathbf{X} = \mathbf{x}\right]\right)\sqrt{\text{Var}\left(Y \vert \mathbf{X}_U = \mathbf{x}'_U, \mathbf{X}_{-U}=\mathbf{X}_{-U}\right)/\text{Var}\left(Y \vert \mathbf{X}=\mathbf{x}\right)} + \\
&\EE\left[Y\vert \mathbf{X}_U = \mathbf{x}'_U, \mathbf{X}_{-U}=\mathbf{X}_{-U}\right]
\end{align*}
for $ \left(\mathbf{X}_U = \mathbf{x}'_U, \mathbf{X}_{-U}=\mathbf{x}_{-U}\right)$ in the support of the observational distribution. Further, $H_{0,U}$ holds, where
\begin{equation}\label{eq:H0j-het}
H_{0,U}: \quad \mathcal{E} \perp \mathbf{X}_U \vert \mathbf{X}_{-U}, \quad \text{with} \quad \mathcal{E} = \dfrac{Y - \EE\left[Y\vert \mathbf{X} \right]}{\sqrt{\text{Var}\left(Y \vert \mathbf{X}\right)}}.
\end{equation}
\end{theo}

By constructing a counterfactual such that the regression residual $\mathcal{E}$ remains unchanged, the effect on $Y$ can be assessed in terms of the conditional mean and the conditional variance. As in Section \ref{local} one could alternatively use do-statements for $ \left(\mathbf{X}_U = \mathbf{x}'_U, \mathbf{X}_{-U}=\mathbf{x}_{-U}\right)$ outside the support of the observational distribution.

\subsection{Asymptotic results}\label{subsec.asymp}
To fit location-scale noise models, a simple approach is to estimate both $\EE\left[Y\vert \mathbf{X}\right]$ and $\EE\left[Y^2\vert \mathbf{X}\right]$. If both these quantities are known, one can recover $\mathcal{E}$.

We consider variations of Algorithms \ref{alg:split} and \ref{alg:multisplit+test} where we get estimates $\hat{f}_1\left(\mathbf{X}\right)$ for $f_1\left(\mathbf{X}\right) \coloneqq \EE\left[Y\vert \mathbf{X}\right]$ and $\hat{f}_2\left(\mathbf{X}\right)$ for $f_2\left(\mathbf{X}\right) \coloneqq\EE\left[Y^2\vert \mathbf{X}\right]$  using certain regressors on the data $\left(\mathbf{x}, \mathbf{y}\right)$; see the notation in Section \ref{estim}. Then, we estimate the residuals
\begin{equation*}
\epsilon_i =\dfrac{y_i - f_1\left(\mathbf{x}_i\right)}{\sqrt{f_2\left(\mathbf{x}_i\right) - f_1^2\left(\mathbf{x}_i\right)}} \quad \text{by} \quad \hat{\epsilon}_i =\dfrac{y_i - \hat{f}_1\left(\mathbf{x}_i\right)}{\sqrt{\hat{f}_2\left(\mathbf{x}_i\right) - \hat{f}_1^2\left(\mathbf{x}_i\right)}}.
\end{equation*}
Especially for low sample sizes, it can happen that $\hat{f}_2\left(\mathbf{x}_i\right)\leq \hat{f}_1^2\left(\mathbf{x}_i\right)$ for some $i$. To make the method operational in such cases, we suggest defining $\hat{\epsilon}_i$ by a large quantity in absolute value with the same sign as $y_i - \hat{f}_1\left(\mathbf{x}_i\right)$. For our asymptotic results, it could even be replaced by arbitrary values.
To establish guarantees for FOCI, we make the following assumptions
\begin{assb}\label{ass:f1}
$\left\vert f_1\left(\mathbf{x}_i\right) - \hat{f}_1\left(\mathbf{x}_i\right) \right\vert = {\scriptstyle \mathcal{O}}_p \left(1\right)$.
\end{assb}
\begin{assb}\label{ass:f2}
$\left\vert f_2\left(\mathbf{x}_i\right) - \hat{f}_2\left(\mathbf{x}_i\right) \right\vert = {\scriptstyle \mathcal{O}}_p \left(1\right)$.
\end{assb}
\begin{assb}\label{ass:var0}
$\PP \left(f_2\left(\mathbf{x}_i\right) - f_1^2\left(\mathbf{x}_i\right)>0\right)=1$.
\end{assb}
In Assumptions \ref{ass:f1} and \ref{ass:f2} the probability is over both, the function estimates and the new data point. Assumption \ref{ass:var0} implies that $Y$ is almost surely not deterministic in $\mathbf{X}$.

\begin{theo}\label{theo:heto-cons}
Suppose that the regularity assumptions \ref{ass:az1} and \ref{ass:az2} \citep{azadkia2021simple} for the data $\left(g\left(\mathcal{E}\right), \mathbf{X}\right)$ hold as well as conditions \ref{ass:inter} and \ref{ass:continuous} - \ref{ass:var0}. Let $\hat{S}$ be the output of Algorithm \ref{alg:split} modified to normalize the residuals for the heteroscedastic noise model. There are positive real numbers $L_1$, $L_2$ and $L_3$ that do not depend on the sample size such that
\begin{equation*}
\PP\left(\hat{S} \supseteq \left\{1,\ldots,p\right\} \setminus W\right) \geq 1 - L_1 p^{L_2} \exp\left(-L_3 n\right).
\end{equation*}
If instead $\hat{S}$ is the output of Algorithm \ref{alg:in} adjusted to normalize the residuals, it holds
\begin{equation*}
\underset{n \rightarrow \infty}{\lim}\PP\left(\hat{S} \supseteq \left\{1,\ldots,p\right\} \setminus W\right) =1.
\end{equation*}
\end{theo}
The key step to adapt the results to the heteroskedastic case is seeing that \ref{ass:f1} - \ref{ass:var0} imply
\begin{equation*}
\left \vert \hat{\epsilon}_i - \epsilon_i \right \vert = {\scriptstyle \mathcal{O}}_p \left(1\right).
\end{equation*}
Then, all the results from the homoskedastic case carry over. Any other regression algorithm tailor-made for location-scale noise models could be applied as well if it ensures this condition.

Although we receive similar asymptotic guarantees for location-scale noise models under rather weak assumptions, they are harder to deal with for finite samples. As all conditional dependence between $Y$ and any $X_j$ that is due to location or scale is regressed out, the residing dependence can be very weak. Hence, the population Conditional Dependence Coefficient \citep{azadkia2021simple} is low requiring an even larger sample size. Also, the absolute value transform appears to be less appropriate after regressing away the scale information. Hence, we apply no transform in the simulation example.

\subsection{Simulation example}
\begin{figure}[h!]
\centering
\begin{tikzpicture}[roundnode/.style={circle, very thick, minimum size=8mm}]
\node[draw, roundnode, text centered] (x1) {${\scriptstyle X_1}$};
\node[draw, roundnode, right =of x1, text centered] (x2) {${\scriptstyle X_2}$};
\node[draw, roundnode, right =of x2, text centered] (y) {${\scriptstyle Y}$};
\node[draw, dotted, roundnode, above right =of x1, text centered] (h) {${\scriptstyle H}$};

\draw[->, line width= 1] (x1) -- (x2);
\draw[->, line width= 1] (x2) -- (y);
\draw[->, line width= 1] (h) -- (x1);
\draw[->, line width= 1] (h) -- (y);

\end{tikzpicture}
 \caption[Sample DAGs]
 {DAG representing the SCM in the simulation fitting LSNM.}
 \label{fig:DAG-sim-het}
\end{figure}
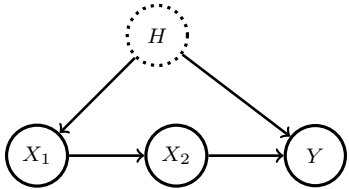

We consider a simple example with two observed predictors and one hidden confounder as shown in Figure \ref{fig:DAG-sim-het}. We let
\begin{equation*}
Y \leftarrow g_{X_2Y}\left(X_2\right)H
\end{equation*}
such that \ref{ass:add-mult-sep} holds for $X_2$. The causal effect is sinusoidal from $H$ to $X_1$, linear from $X_1$ to $X_2$, and there is an additive Gaussian error term on each.

For each sample size from $10^2$ to $10^5$, we run 200 repetitions of the same data generating mechanism. We fit both moments with \texttt{xgboost} and use the identity function for $g\left(\cdot\right)$. Otherwise, we proceed as in Section \ref{sim}.
\begin{figure}[t!]
 \centering
 \includegraphics[width=1\textwidth]{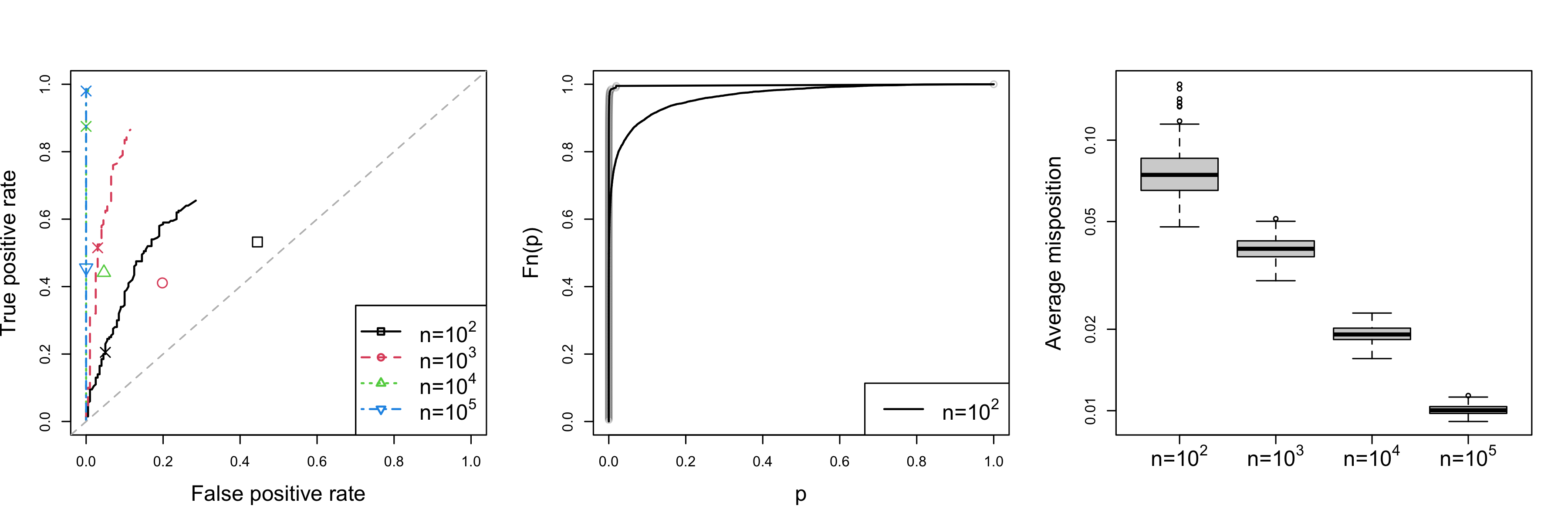}
 \caption[Simulation in a SEM]
 {The results are based on $200$ simulation runs. On the left: False positive rate versus true positive rate obtained with Algorithm \ref{alg:multisplit+test} adjusted to LSNM for varying $\tilde{\alpha}$ and $\alpha = 0.05$. The crosses correspond to $\tilde{\alpha} = 0.01$. The other symbols describe the performance of Algorithm \ref{alg:split} adjusted to LSNM. In the middle: empirical cumulative distribution function of the p-values obtained with HSIC. We compare the raw p-values from each split (lines only) to the cumulated p-value per simulation run (lines with dots). On the right: average misposition \eqref{eq:amp} of the estimated residuals with respect to the true residuals.}
 \label{fig:het-sine}
\end{figure}

In Figure \ref{fig:het-sine} we show the same performance metrics as in Figure \ref{fig:rand-RE}. We see that our method can handle this toy example quite well. For $10^5$ samples, the performance with $\tilde{\alpha}=0.01$ is almost perfect, i.e., $196$ times the output is $\hat{W} = \left\{2\right\}$ and $4$ times $\hat{W} = \emptyset$. There are no false positives in $\hat{W}$.

The global test works well already for $10^2$ samples. After aggregation over the different splits, $H_0$ in \eqref{eq:H0-het} is rejected in every simulation run at $\alpha = 0.05$. This can be facilitated by the fact that the fits are not good for this sample size such that there is more dependence on $\mathbf{x}$ for $\hat{\boldsymbol{\epsilon}}$ than for the true $\boldsymbol{\epsilon}$.

Finally, we compare how well the ordering of $\hat{\boldsymbol{\epsilon}}$ matches that of $\boldsymbol{\epsilon}$. For each run, we calculate the average misposition defined as
\begin{equation}\label{eq:amp}
AMP=\dfrac{1}{n^2}\sum_{i =1}^{n} \left\vert \sum_{l=1}^{n} \mathbbm{1}_{\left\{\epsilon_l < \epsilon_i\right\}} - \mathbbm{1}_{\left\{\hat{\epsilon}_l < \hat{\epsilon}_i\right\}}\right\vert.
\end{equation}
We show the according box plots on the right side of Figure \ref{fig:het-sine}. As desired, this quantity approaches $0$ for increasing sample size. For simplicity, we calculate this quantity only on a single split per simulation run.

\section{Conclusion}
In this paper, we introduce the notion of causal well-specification for additive noise models or their extension to heteroskedastic errors. Our viewpoint of local, i.e., for a subset of the covariates, causal well-specification, for which conditional independence between predictor and residual can serve as a proxy, provides a new option instead of rejecting entire models.

We present an algorithm to estimate our quantities of interest from finite data and provide some asymptotic guarantees. We demonstrate its application in simulation setups. This reveals some difficulties but also shows how considering multiple data splits can help even in hard cases.

Finally, we also apply our methodology and algorithm to regression problems extracted from a large-scale genomic dataset. While in many cases, causal well-specification appears to be not even approximately fulfilled, we find multiple examples where our estimate of well-specification is in line with an approximate validation from various gene knockdown perturbations.

We would like to emphasize that our formulation and analysis of the information provided by conditional independence, which we present in Section \ref{sec:causal-wellspec}, can also be applied as stand-alone and other machine learning methods for regression and conditional dependency assessment can be used.

\subsection*{Acknowledgement}
The project leading to this application has received funding from the European Research Council (ERC) under the European Union’s Horizon 2020 research and innovation programme (grant agreement No 786461).\\
CS thanks Mathieu Chevalley for helpful discussions regarding finding use cases in the K562 dataset.

\bibliographystyle{apalike} 
\bibliography{references_nf}

\begin{thebibliography}{}

\bibitem[Azadkia and Chatterjee, 2021]{azadkia2021simple}
Azadkia, M. and Chatterjee, S. (2021).
\newblock A simple measure of conditional dependence.
\newblock {\em The Annals of Statistics}, 49(6):3070--3102.

\bibitem[Azadkia et~al., 2021]{azadkia2021FOCI}
Azadkia, M., Chatterjee, S., and Matloff, N. (2021).
\newblock {\em FOCI: Feature Ordering by Conditional Independence}.
\newblock R package version 0.1.3.

\bibitem[B{\"u}hlmann et~al., 2014]{buhlmann2014cam}
B{\"u}hlmann, P., Peters, J., and Ernest, J. (2014).
\newblock Cam: Causal additive models, high-dimensional order search and
  penalized regression.
\newblock {\em Ann. Statist.}, 42(6):2526--2556.

\bibitem[Buja et~al., 2019]{buja2019models2}
Buja, A., Brown, L., Kuchibhotla, A.~K., Berk, R., George, E., and Zhao, L.
  (2019).
\newblock {Models as Approximations II: A Model-Free Theory of Parametric
  Regression}.
\newblock {\em Statistical Science}, 34(4):545 -- 565.

\bibitem[Chen et~al., 2021]{chen2021xgboost}
Chen, T., He, T., Benesty, M., Khotilovich, V., Tang, Y., Cho, H., Chen, K.,
  Mitchell, R., Cano, I., Zhou, T., Li, M., Xie, J., Lin, M., Geng, Y., and Li,
  Y. (2021).
\newblock {\em xgboost: Extreme Gradient Boosting}.
\newblock R package version 1.4.1.1.

\bibitem[Chevalley et~al., 2023]{chevalley2022causalbench}
Chevalley, M., Roohani, Y., Mehrjou, A., Leskovec, J., and Schwab, P. (2023).
\newblock Causalbench: A large-scale benchmark for network inference from
  single-cell perturbation data.
\newblock {\em arXiv preprint arXiv:2210.17283}.

\bibitem[Geiger et~al., 1990]{geiger1990identifying}
Geiger, D., Verma, T., and Pearl, J. (1990).
\newblock Identifying independence in bayesian networks.
\newblock {\em Networks}, 20(5):507--534.

\bibitem[Hoyer et~al., 2008]{hoyer2008nonlinear}
Hoyer, P., Janzing, D., Mooij, J.~M., Peters, J., and Sch{\"o}lkopf, B. (2008).
\newblock Nonlinear causal discovery with additive noise models.
\newblock {\em Advances in Neural Information Processing systems}, 21.

\bibitem[Immer et~al., 2023]{immer2022loci}
Immer, A., Schultheiss, C., Vogt, J.~E., Sch{\"o}lkopf, B., B{\"u}hlmann, P.,
  and Marx, A. (2023).
\newblock On the identifiability and estimation of causal location-scale noise
  models.
\newblock In {\em International Conference on Machine Learning}, pages
  14316--14332. PMLR.

\bibitem[Larson et~al., 2013]{larson2013crispr}
Larson, M.~H., Gilbert, L.~A., Wang, X., Lim, W.~A., Weissman, J.~S., and Qi,
  L.~S. (2013).
\newblock Crispr interference (crispri) for sequence-specific control of gene
  expression.
\newblock {\em Nature protocols}, 8(11):2180--2196.

\bibitem[Maeda and Shimizu, 2021]{maeda2021causal}
Maeda, T.~N. and Shimizu, S. (2021).
\newblock Causal additive models with unobserved variables.
\newblock In {\em Uncertainty in Artificial Intelligence}, pages 97--106. PMLR.

\bibitem[McDiarmid et~al., 1989]{mcdiarmid1989method}
McDiarmid, C. et~al. (1989).
\newblock On the method of bounded differences.
\newblock {\em Surveys in Combinatorics}, 141(1):148--188.

\bibitem[Meinshausen and B{\"u}hlmann, 2010]{meinshausen2010stability}
Meinshausen, N. and B{\"u}hlmann, P. (2010).
\newblock Stability selection.
\newblock {\em Journal of the Royal Statistical Society: Series B (Statistical
  Methodology)}, 72(4):417--473.

\bibitem[Meinshausen et~al., 2009]{meinshausen2009p}
Meinshausen, N., Meier, L., and B{\"u}hlmann, P. (2009).
\newblock P-values for high-dimensional regression.
\newblock {\em Journal of the American Statistical Association},
  104(488):1671--1681.

\bibitem[Pearl, 1988]{pearl1988probabilistic}
Pearl, J. (1988).
\newblock {\em Probabilistic reasoning in intelligent systems: networks of
  plausible inference}.
\newblock Morgan Kaufmann Publishers.

\bibitem[Pearl, 2009]{pearl2009causality}
Pearl, J. (2009).
\newblock {\em Causality: Models, Reasoning, and Inference.}
\newblock Cambridge University Press.

\bibitem[Pearl, 2012]{pearl2012calculus}
Pearl, J. (2012).
\newblock The do-calculus revisited.
\newblock In {\em Proceedings of the Twenty-Eighth Conference on Uncertainty in
  Artificial Intelligence}, pages 3--11.

\bibitem[Peters, 2015]{peters2015intersection}
Peters, J. (2015).
\newblock On the intersection property of conditional independence and its
  application to causal discovery.
\newblock {\em Journal of Causal Inference}, 3(1):97--108.

\bibitem[Peters et~al., 2016]{peters2016causal}
Peters, J., B{\"u}hlmann, P., and Meinshausen, N. (2016).
\newblock Causal inference by using invariant prediction: identification and
  confidence intervals.
\newblock {\em Journal of the Royal Statistical Society. Series B (Statistical
  Methodology)}, 78(5):947--1012.

\bibitem[Peters et~al., 2017]{peters2017elements}
Peters, J., Janzing, D., and Sch{\"o}lkopf, B. (2017).
\newblock {\em Elements of causal inference: foundations and learning
  algorithms}.
\newblock The MIT Press.

\bibitem[Peters et~al., 2014]{peters2014causal}
Peters, J., Mooij, J.~M., Janzing, D., and Sch{\"o}lkopf, B. (2014).
\newblock Causal discovery with continuous additive noise models.
\newblock {\em Journal of Machine Learning Research}, 15:2009--2053.

\bibitem[Pfister et~al., 2018]{pfister2018kernel}
Pfister, N., B{\"u}hlmann, P., Sch{\"o}lkopf, B., and Peters, J. (2018).
\newblock Kernel-based tests for joint independence.
\newblock {\em Journal of the Royal Statistical Society. Series B (Statistical
  Methodology)}, 80(1):5--31.

\bibitem[Pfister and Peters, 2019]{pfister2019dHSIC}
Pfister, N. and Peters, J. (2019).
\newblock {\em dHSIC: Independence Testing via Hilbert Schmidt Independence
  Criterion}.
\newblock R package version 2.1.

\bibitem[Replogle et~al., 2022]{replogle2022mapping}
Replogle, J.~M., Saunders, R.~A., Pogson, A.~N., Hussmann, J.~A., Lenail, A.,
  Guna, A., Mascibroda, L., Wagner, E.~J., Adelman, K., Lithwick-Yanai, G.,
  et~al. (2022).
\newblock Mapping information-rich genotype-phenotype landscapes with
  genome-scale perturb-seq.
\newblock {\em Cell}, 185(14):2559--2575.

\bibitem[Rojas-Carulla et~al., 2018]{rojas2018invariant}
Rojas-Carulla, M., Sch{\"o}lkopf, B., Turner, R., and Peters, J. (2018).
\newblock Invariant models for causal transfer learning.
\newblock {\em Journal of Machine Learning Research}, 19(36):1--34.

\bibitem[Schultheiss and B{\"u}hlmann, 2023]{schultheiss2023pitfalls}
Schultheiss, C. and B{\"u}hlmann, P. (2023).
\newblock On the pitfalls of gaussian likelihood scoring for causal discovery.
\newblock {\em Journal of Causal Inference}, 11(1).

\bibitem[Schultheiss et~al., 2023]{schultheiss2023higher}
Schultheiss, C., B{\"u}hlmann, P., and Yuan, M. (2023).
\newblock Higher-order least squares: assessing partial goodness of fit of
  linear causal models.
\newblock {\em Journal of the American Statistical Association}.

\bibitem[Shah and Peters, 2020]{shah2020hardness}
Shah, R.~D. and Peters, J. (2020).
\newblock The hardness of conditional independence testing and the generalised
  covariance measure.
\newblock {\em The Annals of Statistics}, 48(3):1514--1538.

\bibitem[Shah and Samworth, 2013]{shah2013variable}
Shah, R.~D. and Samworth, R.~J. (2013).
\newblock Variable selection with error control: another look at stability
  selection.
\newblock {\em Journal of the Royal Statistical Society: Series B (Statistical
  Methodology)}, 75(1):55--80.

\bibitem[Spirtes, 2001]{spirtes2001anytime}
Spirtes, P. (2001).
\newblock An anytime algorithm for causal inference.
\newblock In {\em International Workshop on Artificial Intelligence and
  Statistics}, pages 278--285. PMLR.

\bibitem[Spirtes et~al., 2000]{spirtes2000causation}
Spirtes, P., Glymour, C.~N., and Scheines, R. (2000).
\newblock {\em Causation, prediction, and search}.
\newblock MIT press.

\bibitem[Strobl and Lasko, 2023]{strobl2022identifying}
Strobl, E.~V. and Lasko, T.~A. (2023).
\newblock Identifying patient-specific root causes with the heteroscedastic
  noise model.
\newblock {\em Journal of Computational Science}, 72.

\bibitem[Wang et~al., 2023]{wang2022generalizing}
Wang, J., Lan, C., Liu, C., Ouyang, Y., Qin, T., Lu, W., Chen, Y., Zeng, W.,
  and Yu, P. (2023).
\newblock Generalizing to unseen domains: A survey on domain generalization.
\newblock {\em IEEE Transactions on Knowledge and Data Engineering},
  35(8):8052--8072.

\bibitem[Wood, 2011]{wood2011fast}
Wood, S.~N. (2011).
\newblock Fast stable restricted maximum likelihood and marginal likelihood
  estimation of semiparametric generalized linear models.
\newblock {\em Journal of the Royal Statistical Society (B)}, 73(1):3--36.

\bibitem[Xu et~al., 2022]{xu2022inferring}
Xu, S., Mian, O.~A., Marx, A., and Vreeken, J. (2022).
\newblock Inferring cause and effect in the presence of heteroscedastic noise.
\newblock In {\em International Conference on Machine Learning}, pages
  24615--24630. PMLR.

\end{thebibliography}

\newpage
\appendix
\allowdisplaybreaks
\section{Proofs}
\subsection{Proof of Theorem \ref{theo:H0j}}\label{proof:H0j}
Recall
\begin{equation*}
Y \coloneqq  \EE\left[Y \vert \mathbf{X}\right] + \mathcal{E}
\end{equation*}
Due to \ref{ass:add-sep}, we have
\begin{align*}
\EE\left[Y \vert \mathbf{X}\right] &=  f_{\mathbf{X}_U Y}\left(\mathbf{X}_U, \mathbf{X}_{\text{PA}\left(Y\right) \setminus U}\right) + \EE\left[f_{\mathbf{H}Y}\left(\mathbf{H}_{\text{PA}\left(Y\right)}, \mathbf{X}_{\text{PA}\left(Y\right) \setminus U}\right) \vert \mathbf{X}\right]  \quad \text{such that}\\
\mathcal{E} &=f_{\mathbf{H}Y}\left(\mathbf{H}_{\text{PA}\left(Y\right)}, \mathbf{X}_{\text{PA}\left(Y\right) \setminus U}\right) - \EE\left[f_{\mathbf{H}Y}\left(\mathbf{H}_{\text{PA}\left(Y\right)}, \mathbf{X}_{\text{PA}\left(Y\right) \setminus U}\right) \vert \mathbf{X}\right].
\end{align*}
Using \ref{ass:markov}, $\mathbf{X}_U \perp \mathbf{H}_{\text{PA}\left(Y\right)} \vert \mathbf{X}_{-U}$, and trivially, $\mathbf{X}_U \perp \mathbf{X}_{-U} \vert \mathbf{X}_{-U}$. It follows 
\begin{align*}
\EE\left[f_{\mathbf{H}Y}\left(\mathbf{H}_{\text{PA}\left(Y\right)}, \mathbf{X}_{\text{PA}\left(Y\right) \setminus U}\right) \vert \mathbf{X}\right] = \EE\left[f_{\mathbf{H}Y}\left(\mathbf{H}_{\text{PA}\left(Y\right)}, \mathbf{X}_{\text{PA}\left(Y\right) \setminus U}\right) \vert \mathbf{X}_{-U}\right] & \perp \mathbf{X}_U \vert \mathbf{X}_{-U} \quad \text{and} \\
f_{\mathbf{H}Y}\left(\mathbf{H}_{\text{PA}\left(Y\right)}, \mathbf{X}_{\text{PA}\left(Y\right)\setminus j}\right)  \perp \mathbf{X}_U \vert \mathbf{X}_{-U} \quad \text{such that} \quad \mathcal{E} & \perp \mathbf{X}_U  \vert \mathbf{X}_{-U}.
\end{align*}
Consider the counterfactual intervention. As $\mathbf{X}_{-U}$ remains unchanged, the second summand in \ref{ass:add-sep} could only change if $\mathbf{H}_{\text{PA}\left(Y\right)}$ changes. This could happen through some directed path from $\mathbf{X}_U$ to $\mathbf{H}_{\text{PA}\left(Y\right)}$ that is not blocked by $\mathbf{X}_{-U}$. By \ref{ass:markov}, if such an effect from $\mathbf{X}_U$ to $\mathbf{H}_{\text{PA}\left(Y\right)}$ exists, it is constant for almost all $\mathbf{x}_U$. With \ref{ass:indep}, we can extend this argument to all attainable $\mathbf{x}_U$. Hence, changing $\mathbf{X}_U$ from $\mathbf{x}_U$ to $\mathbf{x}'_U$ while keeping $\mathbf{X}_{-U}$ fixed, cannot affect $\mathbf{H}_{\text{PA}\left(Y\right)}$ such that the second summand remains constant. For the first summand, we can directly plug in the counterfactual values of $\mathbf{X}$.

In the conditional expectation given above only the first summand can change as the second is a function of only $\mathbf{X}_{-U}$. As the altered summand is the same for both $Y$ and $\EE\left[Y \vert \mathbf{X}\right]$, the new value $y'$ must exactly represent this change in conditional mean.

\subsection{Proof of Theorem \ref{theo:WW}}
Consider first the $\subseteq$-statement. This means that $H_{0,j}$ in \eqref{eq:H0j-ind} must hold $\forall j \in W$. Let $S = \left\{1,\ldots,p\right\} \setminus W$. Then, we want that
\begin{equation*}
\mathcal{E} \perp \mathbf{X}_W \vert \mathbf{X}_S \implies \mathcal{E} \perp X_j \vert \mathbf{X}_{-j}. 
\end{equation*}
This can be rewritten as
\begin{equation*}
\mathcal{E} \perp \mathbf{X}_{W \setminus j}, X_j \vert \mathbf{X}_S \implies \mathcal{E} \perp X_j \vert \mathbf{X}_S, \mathbf{X}_{W \setminus j}. 
\end{equation*}
This is the weak union property in Chapter 3 of \cite{pearl1988probabilistic} and hence holds for any random variables. 

For $W = \tilde{W}$, we additionally need that $H_{0,j}$ cannot hold for any  $j \in S$. By minimality of $S$
\begin{equation*}
\mathcal{E} \not \perp X_j, \mathbf{X}_W \vert \mathbf{X}_{S \setminus j}.
\end{equation*}
Then, the intersection property implies
\begin{equation*}
\mathcal{E} \not \perp \mathbf{X}_W \vert X_j, \mathbf{X}_{S \setminus j} \quad \text{or} \quad \mathcal{E} \not \perp X_j \vert \mathbf{X}_W, \mathbf{X}_{S \setminus j}.
\end{equation*}
The first cannot hold by the definition of $W$, so the second must hold. This means that $H_{0,j}$ is not fulfilled, and $W= \tilde{W}$ is guaranteed. As $\tilde{W}$ is unique by construction, $W$ is then unique as well.

\subsection{Definitions from FOCI}\label{app:FOCI}
These definitions are taken from \citep{azadkia2021simple}, adapted in parts to fit our notation. Let $\mu$ be the law of $\mathcal{E}$. We have the following population quantities
\begin{align*}
Q\left(\mathcal{E}, \mathbf{X}_U\right) &= \int\text{Var}\left(\PP\left( \mathcal{E}\geq t \vert \mathbf{X}_U\right)\right) d\mu\left(t\right) \geq 0 \\
S\left(\mathcal{E} \right) & = \int\text{Var}\left(\mathbbm{1}_{\mathcal{E}\geq t}\right) d\mu\left(t\right) \geq 0 \\
T\left(\mathcal{E}, \mathbf{X}_U\right) & = Q\left(\mathcal{E}, \mathbf{X}_U\right) / S\left(\mathcal{E} \right) \in \left[0,1\right]\\
Q\left(\mathcal{E}, \mathbf{X}_U \vert \mathbf{X}_S\right) &= \int\EE\left[\text{Var}\left(\PP\left( \mathcal{E}\geq t \vert \mathbf{X}_U, \mathbf{X}_S\right) \vert \mathbf{X}_S\right)\right] d\mu\left(t\right) \geq 0 \\
S\left(\mathcal{E}, \mathbf{X}_S \right) & =\int \EE\left[\text{Var}\left(\mathbbm{1}_{\mathcal{E}\geq t} \vert \mathbf{X}_S\right)\right] d\mu\left(t\right) \geq 0\\
T\left(\mathcal{E}, \mathbf{X}_U \vert \mathbf{X}_S\right) & = Q\left(\mathcal{E}, \mathbf{X}_U \vert \mathbf{X}_S\right) / S\left(\mathcal{E}, \mathbf{X}_S \right) \in \left[0,1\right].
\end{align*}
For data estimates, define first
\begin{equation*}
R_i = \sum_{l=1}^n \mathbbm{1}_{\epsilon_l \leq \epsilon_i}, \quad L_i = \sum_{l=1}^n \mathbbm{1}_{\epsilon_l \geq \epsilon_i}
\end{equation*}
and $M\left(i\right)$ the nearest neighbour of $i$ with respect to $\mathbf{x}_U$ with a random tie-breaking rule. Then, we have the data estimates
\begin{align*}
Q_n\left(\boldsymbol{\epsilon}, \mathbf{x}_U\right) &= \dfrac{1}{n^2} \sum_{i=1}^n \min\left\{R_i, R_{M\left(i\right)}\right\} - \dfrac{L_i^2}{n} \\
S_n \left(\boldsymbol{\epsilon}\right) &= \dfrac{1}{n^3} \sum_{i=1}^n L_i\left(n - L_i\right)\\
T_n\left(\boldsymbol{\epsilon}, \mathbf{x}_U\right) &= Q_n\left(\boldsymbol{\epsilon}, \mathbf{x}_U\right) / S_n \left(\boldsymbol{\epsilon}\right)\\
S_n \left(\boldsymbol{\epsilon}, \mathbf{x}_U\right) &= \dfrac{1}{n^2} \sum_{i=1}^n R_i - \min\left\{R_i, R_{M\left(i\right)}\right\}.\\
\end{align*}

\subsection{Proof of Proposition \ref{prop:abs}}
We have $T\left(\cdot, \mathbf{X}_S\right) = Q\left(\cdot, \mathbf{X}_S\right)/S\left(\cdot\right)$.
As argued in \citep{azadkia2021simple} the denominator for unconditional independence tests is simply $1/6$ for continuous random variables.
If $\mathcal{E}$ is conditionally continuously distributed, the same holds for its marginal distribution and thus also for the distribution of $\left\vert \mathcal{E}\right\vert$. Hence, it suffices to consider $Q\left(\cdot, \mathbf{X}_S\right)$ and the statement for $T\left(\cdot, \mathbf{X}_S\right)$ follows directly. Let $\mu$ and $\nu$ be the law of $\mathcal{E}$ and  $\left\vert \mathcal{E}\right\vert$. Due to symmetry, it holds
$d\nu\left(t\right) = 2 d\mu\left(t\right) \ \forall t \geq 0$.
\begin{align*}
Q\left(\left\vert \mathcal{E}\right\vert, \mathbf{X}_S\right) = &\int_{0}^{\infty}\text{Var}\left(\PP\left(\left\vert \mathcal{E}\right\vert\geq t \vert \mathbf{X}_S\right)\right) d\nu\left(t\right) = \int_{0}^{\infty}\text{Var}\left(2\PP\left( \mathcal{E}\geq t \vert \mathbf{X}_S\right)\right) 2 d\mu\left(t\right)\\
 = 8 &\int_{0}^{\infty}\text{Var}\left(\PP\left( \mathcal{E}\geq t \vert \mathbf{X}_S\right)\right) d\mu\left(t\right) \\
 &\int_{-\infty}^{0}\text{Var}\left(\PP\left( \mathcal{E}\geq t \vert \mathbf{X}_S\right)\right) d\mu\left(t\right) = \int_{-\infty}^{0}\text{Var}\left(\PP\left( \mathcal{E}\leq -t \vert \mathbf{X}_S\right)\right) d\mu\left(t\right) \\
=&\int_{-\infty}^{0}\text{Var}\left(1-\PP\left( \mathcal{E}\geq -t \vert \mathbf{X}_S\right)\right) d\mu\left(t\right) = \int_{-\infty}^{0}\text{Var}\left(\PP\left( \mathcal{E}\geq -t \vert \mathbf{X}_S\right)\right) d\mu\left(t\right) \\
 \overset{t' \leftarrow -t}{=} &\int_{0}^{\infty}\text{Var}\left(\PP\left( \mathcal{E}\geq t' \vert \mathbf{X}_S\right)\right) d\mu\left(t'\right) \ \text{such that} \\
Q\left(\mathcal{E}, \mathbf{X}_S\right) =& \int_{-\infty}^{\infty}\text{Var}\left(\PP\left( \mathcal{E}\geq t \vert \mathbf{X}_S\right)\right) d\mu\left(t\right) = 2\int_{0}^{\infty}\text{Var}\left(\PP\left( \mathcal{E}\geq t \vert \mathbf{X}_S\right)\right) d\mu\left(t\right).
\end{align*}
The first line uses symmetry, and the second chain of equalities uses symmetry as well as continuity to allow for a weak inequality in the complementary probability.
Comparing the quantity on the first line to that on the last line we see that the ratio between the numerator terms is $4$.

\subsection{Proof of Theorem \ref{theo:cons}}
We build up the proof by some supporting Lemmata.
\begin{lemm}\label{lemm:prob}
Assume \ref{ass:suitable} and \ref{ass:continuous}.
\begin{equation*}
\underset{n \rightarrow \infty}{\lim} \PP\left(\left[g\left(\epsilon_i\right) > g\left(\epsilon_l\right) \cap g\left(\hat{\epsilon}_i\right) \leq g\left(\hat{\epsilon}_l\right)\right] \cup \left[g\left(\epsilon_i\right) < g\left(\epsilon_l\right) \cap g\left(\hat{\epsilon}_i\right) \geq g\left(\hat{\epsilon}_l\right)\right]\cup \left[g\left(\epsilon_i\right) = g\left(\epsilon_l\right)\right]\right)=0 \ \forall i \neq l,
\end{equation*}
i.e., the probability that the estimates imply a different ordering between $i$ and $l$ approaches $0$.
\end{lemm}

Define $Q_n\left(\cdot\right)$ and $S_n\left(\cdot\right)$ as in Section 9 of \cite{azadkia2021simple}.

\begin{lemm}\label{lemm:mean}
Assume the conditions of Lemma \ref{lemm:prob}. Let $U$ be any non-empty subset of $\left\{1,\ldots,p\right\}$. Then,
\begin{align*}
\underset{n \rightarrow \infty}{\lim}&\EE\left[\left\vert Q_n\left(g\left(\boldsymbol{\epsilon}\right),\mathbf{x}_U\right)-Q_n\left(g\left(\hat{\boldsymbol{\epsilon}}\right),\mathbf{x}_U\right)\right\vert\right]=0, \ \underset{n \rightarrow \infty}{\lim}\EE\left[\left\vert 
S_n\left(g\left(\boldsymbol{\epsilon}\right),\mathbf{x}_U\right)-S_n\left(g\left(\hat{\boldsymbol{\epsilon}}\right),\mathbf{x}_U\right)\right\vert\right]=0 \\
\underset{n \rightarrow \infty}{\lim}&\EE\left[\left\vert 
S_n\left(g\left(\boldsymbol{\epsilon}\right)\right)-S_n\left(g\left(\hat{\boldsymbol{\epsilon}}\right)\right)\right\vert\right]=0.
\end{align*}
\end{lemm}

As in the sample splitting case $\left(g\left(\hat{\boldsymbol{\epsilon}}\right),\mathbf{x}_U\right)$ are i.i.d.\ copies, one can apply Lemma 11.9 in \cite{azadkia2021simple} to those. This yields
\begin{equation}\label{eq:exp}
\PP \left(\left\vert Q_n\left(g\left(\hat{\boldsymbol{\epsilon}}\right),\mathbf{x}_U\right) - \EE\left[Q_n\left(g\left(\hat{\boldsymbol{\epsilon}}\right),\mathbf{x}_U\right)\right]\right\vert \geq t \right)\leq K_1 \exp\left(-K_2 nt^2\right),
\end{equation}
for some positive $K_1$, $K_2$. Therefore, we can draw similar conclusions as in their Lemma 14.2.
\begin{lemm}\label{lemm:exp}
Let $U$ be a subset of size $u$. Assume conditions (A1), which defines $\beta$, and (A2) from \cite{azadkia2021simple} for the data $\left(g\left(\mathcal{E}\right), \mathbf{X}_U\right)$ as well as conditions \ref{ass:suitable} - \ref{ass:continuous}. Then, there exist positive $K_1$, $K_2$, and $K_3$ that do not depend on the sample size such that in the sample splitting case
\begin{align*}
&\PP\left(\left\vert Q_n\left(g\left(\hat{\boldsymbol{\epsilon}}\right),\mathbf{x}_U\right) - Q\left(g\left(\mathcal{E}\right),\mathbf{X}_U\right)\right\vert \geq K_1 \max\left\{D^{1/3}\left(
n\right),n^{-\min\left\{-1/u,-1/2\right\}}\log\left(n\right)^{u+\beta + 1}\right\} + t \right) \leq \\
& K_2 \exp\left(-K_3 nt^2\right)
\end{align*}
\end{lemm}

Under \ref{ass:inter} and \ref{ass:dep} any set $U$ that is not a (weak) superset of $\left\{1,\ldots,p\right\} \setminus W$ cannot be sufficient for $g\left(\mathcal{E}\right)$. Thus, it suffices to bound the probability of $\hat{S}$ not being sufficient, and then Theorem \ref{theo:cons} follows. This corresponds to Theorem 6.1 in \cite{azadkia2021simple}. The only part of its proof that needs adaptation is Lemma 16.3. To proof an according result based on our Lemma \ref{lemm:exp}, we require
\begin{equation*}
L_1 \max\left\{D\left(
n\right),n^{-\min\left\{-1/K,-1/2\right\}}\log\left(n\right)^{K+\beta + 1}\right\} \leq \dfrac{\delta}{16}.
\end{equation*}
Here, we use their definition of $\delta$, i.e., $\delta$ is the largest number such that for any insufficient subset $U \not \supseteq \left(\left\{1,\ldots,p\right\} \setminus W\right)$, there exists $j \not \in U$ that fulfils $Q\left(g\left(\mathcal{E}\right),\mathbf{X}_{U\cup j}\right) \geq Q\left(g\left(\mathcal{E}\right),\mathbf{X}_U\right) + \delta$. $K$ is the integer part of $1/\delta + 2$. As we consider fixed data generating mechanisms, $\delta > 0$ holds by construction. Hence, we do not mention it in the theorems explicitly.
This inequality might require a larger sample size than in \cite{azadkia2021simple} and larger $L_6$ accordingly. Apart from that, the proof follows from the same principles.

\subsubsection{Proof of Lemma \ref{lemm:prob}}
The properties of $g\left(\cdot\right)$ imply that $g\left(\mathcal{E}\right)$ is a continuous random variable as well such that the probability of the last event has probability $0$ regardless of the sample size. As $i$ and $l$ are interchangeable, the first two events have the same probability and it suffices to analyse one. Let $\eta > 0$ be arbitrary.
\begin{align*}
& \PP\left(g\left(\epsilon_i\right) > g\left(\epsilon_l\right) \cap g\left(\hat{\epsilon}_i\right) \leq g\left(\hat{\epsilon}_l\right)\right)=\\
& \PP\left(g\left(\epsilon_i\right) > g\left(\epsilon_l\right) \cap g\left(\hat{\epsilon}_i\right) \leq g\left(\hat{\epsilon}_l\right)\cap g\left(\epsilon_i\right) - g\left(\epsilon_l\right)\leq \eta \right) +\\
& \PP\left(g\left(\epsilon_i\right) > g\left(\epsilon_l\right) \cap g\left(\hat{\epsilon}_i\right)\leq g\left(\hat{\epsilon}_l\right)\cap g\left(\epsilon_i\right) - g\left(\epsilon_l\right) > \eta \right) \leq \\ 
&\PP\left( \left\vert g\left(\epsilon_i\right) - g\left(\epsilon_l\right) \right \vert \leq \eta \right)+\PP\left(\left\vert g\left(\hat{\epsilon}_i\right)-g\left(\epsilon_i\right)\right\vert +  \left\vert g\left(\hat{\epsilon}_l\right)-g\left(\epsilon_l\right)\right\vert  > \eta \right) \leq \\
&\PP\left( \left\vert g\left(\epsilon_i\right) - g\left(\epsilon_l\right) \right \vert \leq \eta \right)+\PP\left(\max\left\{\left\vert g\left(\hat{\epsilon}_i\right)-g\left(\epsilon_i\right)\right\vert , \left\vert g\left(\hat{\epsilon}_l\right)-g\left(\epsilon_l\right)\right\vert \right\}  > \eta/2 \right)\leq \\
&\PP\left( \left\vert g\left(\epsilon_i\right) - g\left(\epsilon_l\right) \right \vert \leq \eta \right)+2\PP\left(\left\vert g\left(\hat{\epsilon}_i\right)-g\left(\epsilon_i\right)\right\vert > \eta/2 \right)\leq \PP\left( \left\vert g\left(\epsilon_i\right) - g\left(\epsilon_l\right) \right \vert \leq \eta \right)+2\PP\left(\left\vert \hat{\epsilon}_i-\epsilon_i\right\vert > \eta/2l \right)
\end{align*}
Let now $\eta$ depend on $n$. For $\eta \rightarrow 0$ the first term vanishes. If it approaches $0$ slowly enough, the second term vanishes as well assuming the regression is suitable. Thus, one can choose $\eta$ such that both terms vanish. Since the inequality holds for arbitrary $\eta$, the probability goes to $0$, i.e.,
\begin{equation*}
\PP \left(\left[g\left(\epsilon_l\right) \leq g\left(\epsilon_i\right) \cap g\left(\hat{\epsilon}_l\right) > g\left(\hat{\epsilon}_i\right)\right]\cup \left[g\left(\epsilon_l\right) \geq g\left(\epsilon_i\right) \cap g\left(\hat{\epsilon}_l\right) < g\left(\hat{\epsilon}_i\right)\right] \right) = {\scriptstyle \mathcal{O}}\left(1\right).
\end{equation*}

\subsubsection{Proof of Lemma \ref{lemm:mean}}
Let $R_i=\sum g\left(\epsilon_l\right) \leq g\left(\epsilon_i\right)$, $L_i= \sum g\left(\epsilon_l\right) \geq g\left(\epsilon_i\right)$, and $\hat{R}_i$, $\hat{L}_i$ the according quantities estimated with $\hat{\boldsymbol{\epsilon}}$.
Note that index $M\left(i\right)$, i.e., the nearest neighbour of $i$ with respect to $\mathbf{x}_U$, only depends on observed quantities. Hence, it is the same for the estimated quantity $\hat{R}_{M\left(i\right)}$.
\begin{align*}
&\left\vert Q_n\left(g\left(\boldsymbol{\epsilon}\right),\mathbf{x}_U\right)-Q_n\left(g\left(\hat{\boldsymbol{\epsilon}}\right),\mathbf{x}_U\right)\right\vert = \left\vert \dfrac{1}{n^2} \sum_{i=1}^n \min\left\{R_i, R_{M\left(i\right)}\right\} - \min\left\{\hat{R}_i, \hat{R}_{M\left(i\right)}\right\} + \dfrac{\hat{L}_i^2-L_i^2}{n} \right\vert \leq \\
& \left\vert \dfrac{1}{n^2} \sum_{i=1}^n \min\left\{R_i, R_{M\left(i\right)}\right\} - \min\left\{\hat{R}_i, \hat{R}_{M\left(i\right)}\right\}\right\vert + \left\vert \dfrac{1}{n^3} \sum_{i=1}^n \hat{L}_i^2-L_i^2 \right\vert \\
&\left\vert S_n\left(g\left(\boldsymbol{\epsilon}\right),\mathbf{x}_U\right)-S_n\left(g\left(\hat{\boldsymbol{\epsilon}}\right),\mathbf{x}_U\right)\right\vert =\left\vert \dfrac{1}{n^2} \sum_{i=1}^n R_i -\hat{R}_i + \min\left\{\hat{R}_i, \hat{R}_{M\left(i\right)}\right\}-\min\left\{R_i, R_{M\left(i\right)}\right\} \right\vert \leq \\
&\left\vert \dfrac{1}{n^2} \sum_{i=1}^n R_i -\hat{R}_i \right\vert + \left\vert \dfrac{1}{n^2} \sum_{i=1}^n \min\left\{\hat{R}_i, \hat{R}_{M\left(i\right)}\right\}-\min\left\{R_i, R_{M\left(i\right)}\right\} \right\vert \\
& \left\vert S_n\left(g\left(\boldsymbol{\epsilon}\right)\right)-S_n\left(g\left(\hat{\boldsymbol{\epsilon}}\right)\right)\right\vert = \left\vert \dfrac{1}{n^3} \sum_{i=1}^n n\left(L_i -\hat{L}_i\right) + \hat{L}_i^2 - L_i^2 \right\vert \leq \left\vert \dfrac{1}{n^2} \sum_{i=1}^n L_i -\hat{L}_i \right\vert + \left\vert \dfrac{1}{n^3} \sum_{i=1}^n \hat{L}_i^2 - L_i^2 \right\vert
\end{align*}
Thus, there are four different terms to be controlled. If both $\boldsymbol{\epsilon}$ and $\hat{\boldsymbol{\epsilon}}$ have $n$ distinct values, all the terms that do not depend on the nearest neighbouring property amongst $\mathbf{x}_U$ are trivially $0$ for all sample sizes. However, we can prove convergence without this assumption.
\begin{align*}
&\EE\left[\left\vert \dfrac{1}{n^2} \sum_{i=1}^n R_i -\hat{R}_i \right\vert\right] \leq \EE\left[ \dfrac{1}{n^2} \sum_{i=1}^n \left\vert R_i -\hat{R}_i \right\vert\right] = \EE\left[ \dfrac{1}{n^2} \sum_{i=1}^n \left\vert \sum_{l=1}^n \mathbbm{1}_{\left\{g\left(\epsilon_l\right) \leq g\left(\epsilon_i\right)\right\}} - \mathbbm{1}_{\left\{g\left(\hat{\epsilon}_l\right) \leq g\left(\hat{\epsilon}_i\right)\right\}} \right\vert\right]=\\
& \EE\left[ \dfrac{1}{n^2} \sum_{i=1}^n \left\vert \sum_{l\neq i} \mathbbm{1}_{\left\{g\left(\epsilon_l\right) \leq g\left(\epsilon_i\right)\right\}} - \mathbbm{1}_{\left\{g\left(\hat{\epsilon}_l\right) \leq g\left(\hat{\epsilon}_i\right)\right\}} \right\vert\right] \leq \dfrac{1}{n^2} \sum_{i=1}^n \sum_{l \neq i} \EE\left[\left\vert \mathbbm{1}_{\left\{g\left(\epsilon_l\right) \leq g\left(\epsilon_i\right)\right\}} - \mathbbm{1}_{\left\{g\left(\hat{\epsilon}_l\right) \leq g\left(\hat{\epsilon}_i\right)\right\}} \right\vert\right] =\\
& \dfrac{n^2-n}{n^2}\EE\left[\left\vert \mathbbm{1}_{\left\{g\left(\epsilon_l\right) \leq g\left(\epsilon_i\right)\right\}} - \mathbbm{1}_{\left\{g\left(\hat{\epsilon}_l\right) \leq g\left(\hat{\epsilon}_i\right)\right\}} \right\vert\right]=\\
&\dfrac{n^2-n}{n^2}\PP \left(\left[g\left(\epsilon_l\right) \leq g\left(\epsilon_i\right) \cap g\left(\hat{\epsilon}_l\right) > g\left(\hat{\epsilon}_i\right)\right]\cup \left[g\left(\epsilon_l\right) \geq g\left(\epsilon_i\right) \cap g\left(\hat{\epsilon}_l\right) < g\left(\hat{\epsilon}_i\right)\right] \right)\overset{n\rightarrow \infty}{\rightarrow}0
\end{align*}
by Lemma \ref{lemm:prob}. In the last two expressions, $l\neq i$ is assumed. The argument for the term with $L_i - \hat{L}_i$ is identical.
\begin{align*}
&\EE\left[\left\vert \dfrac{1}{n^2} \sum_{i=1}^n \min\left\{\hat{R}_i, \hat{R}_{M\left(i\right)}\right\}-\min\left\{R_i, R_{M\left(i\right)}\right\} \right\vert \leq \EE\left[\dfrac{1}{n^2} \sum_{i=1}^n \left \vert \hat{R}_i - R_i\right\vert + \left \vert \hat{R}_{M\left(i\right)} - R_{M\left(i\right)}\right\vert\right]\right] = \\
& \EE\left[\dfrac{1}{n^2} \sum_{i=1}^n  \left \vert \hat{R}_i - R_i\right\vert + \dfrac{1}{n^2} \sum_{i=1}^n \sum_{l: \ M\left(l\right)=i}\left \vert \hat{R}_i - R_i\right\vert\right] =\EE\left[\dfrac{1}{n^2} \sum_{i=1}^n \left \vert \hat{R}_i - R_i\right\vert \left(1 + \sum_{l \neq i} \mathbbm{1}_{M\left(l\right)=i}\right)\right]=\\
&\dfrac{1}{n^2} \sum_{i=1}^n \EE\left[\left\vert \hat{R}_i - R_i\right\vert \left(1 + \sum_{l \neq i} \mathbbm{1}_{M\left(l\right)=i}\right)\right] =  \dfrac{1}{n^2} \sum_{i=1}^n \EE\left[\left\vert \hat{R}_i - R_i\right\vert \EE\left[1 + \sum_{l \neq i} \mathbbm{1}_{M\left(l\right)=i}\vert \hat{R}_i , R_i\right]\right] \leq \\
& \dfrac{2 + C\left(p\right)}{n^2} \sum_{i=1}^n \EE\left[\left \vert \hat{R}_i - R_i\right\vert\right]\overset{n\rightarrow \infty}{\rightarrow}0.
\end{align*}
By Lemma 11.4 in \cite{azadkia2021simple}, there is a dimension-dependent constant such that no point can be the nearest neighbour of more than $C\left(p\right)$ distinct points in $\mathbb{R}^p$. If there are $l$ such that $\mathbf{x}_{l,U}=\mathbf{x}_{i,U}$, $M\left(l\right)$ is chosen uniformly at random from this set, and in expectation there is one $l$ such that $M\left(l\right)=i$. As this uniform draw is independent of $R_i$ and $\hat{R}_i$, the upper bound also applies to the conditional expectation and we can pull it out.
\begin{align*}
&\EE\left[\left\vert \dfrac{1}{n^3} \sum_{i=1}^n L_i^2 -\hat{L}_i^2 \right\vert\right]=\EE\left[\left\vert \dfrac{1}{n^3} \sum_{i=1}^n \left(L_i -\hat{L}_i\right)\left(L_i +\hat{L}_i\right) \right\vert\right]\leq \EE\left[\left\vert \dfrac{2}{n^2} \sum_{i=1}^n L_i -\hat{L}_i \right\vert\right]\overset{n\rightarrow \infty}{\rightarrow}0.
\end{align*}
Thus, every term is under control which concludes the proof. As all the terms are at most of the same order as the probability in Lemma \ref{lemm:prob}, the bound on the convergence rate follows directly.

\subsubsection{Proof of Lemma \ref{lemm:exp}}
By Lemma \ref{lemm:mean}, there exists a rate, say, $D\left(n\right)={\scriptstyle \mathcal{O}}\left(1\right)$ such that 
\begin{equation*}
\left\vert \EE\left[Q_n\left(g\left(\hat{\boldsymbol{\epsilon}}\right),\mathbf{x}_U\right)\right] -\EE\left[Q_n\left(g\left(\epsilon\right),\mathbf{x}_U\right)\right] \right\vert = \mathcal{O}\left(D\left(n\right)\right).
\end{equation*}
Then,
\begin{align*}
& \left\vert Q_n\left(g\left(\hat{\boldsymbol{\epsilon}}\right),\mathbf{x}_U\right) - Q\left(g\left(\mathcal{E}\right),\mathbf{X}_U\right)\right\vert \leq \\
&\left\vert Q_n\left(g\left(\hat{\boldsymbol{\epsilon}}\right),\mathbf{x}_U\right) - \EE\left[Q_n\left(g\left(\hat{\boldsymbol{\epsilon}}\right),\mathbf{x}_U\right)\right]\right\vert +\left\vert \EE\left[Q_n\left(g\left(\hat{\boldsymbol{\epsilon}}\right),\mathbf{x}_U\right)\right] -\EE\left[Q_n\left(g\left(\epsilon\right),\mathbf{x}_U\right)\right] \right\vert+ \\
&\left\vert \EE\left[Q_n\left(g\left(\epsilon\right),\mathbf{x}_U\right)- Q\left(g\left(\mathcal{E}\right),\mathbf{X}_U\right)\right] \right\vert \leq \\
&\left\vert Q_n\left(g\left(\hat{\boldsymbol{\epsilon}}\right),\mathbf{x}_U\right) - \EE\left[Q_n\left(g\left(\hat{\boldsymbol{\epsilon}}\right),\mathbf{x}_U\right)\right]\right\vert + K_1 \max\left\{D\left(
n\right),n^{-\min\left\{-1/u,-1/2\right\}}\log\left(n\right)^{u+\beta + 1}\right\},
\end{align*}
using the rate derived in Lemma 14.2 of \citep{azadkia2021simple}. Therefore, with \eqref{eq:exp},
\begin{align*}
&\PP\left(\left\vert Q_n\left(g\left(\hat{\boldsymbol{\epsilon}}\right),\mathbf{x}_U\right) - Q\left(g\left(\mathcal{E}\right),\mathbf{X}_U\right)\right\vert \geq K_1 \max\left\{D\left(
n\right),n^{-\min\left\{-1/u,-1/2\right\}}\log\left(n\right)^{u+\beta + 1}\right\} + t \right) \leq \\
&\PP \left(\left\vert Q_n\left(g\left(\hat{\boldsymbol{\epsilon}}\right),\mathbf{x}_U\right) - \EE\left[Q_n\left(g\left(\hat{\boldsymbol{\epsilon}}\right),\mathbf{x}_U\right)\right]\right\vert \geq t \right)\leq K_2 \exp\left(-K_3 nt^2\right).
\end{align*}

\subsection{Proof of Theorem \ref{theo:cons2}}
Again, we only have to bound the probability of $\hat{S}$ not being sufficient.

Using Lemma \ref{lemm:mean} and the Markov inequality, we see
\begin{equation*}
\PP \left(\left\vert Q_n\left(g\left(\hat{\boldsymbol{\epsilon}}\right),\mathbf{x}_U\right) - Q_n\left(g\left(\boldsymbol{\epsilon}\right),\mathbf{x}_U\right)\right\vert \geq t \right)\leq \dfrac{K_1 D\left(n\right)}{t}.
\end{equation*}
Hence, we get
\begin{align*}
\PP&\left(\left\vert Q_n\left(g\left(\hat{\boldsymbol{\epsilon}}\right),\mathbf{x}_U\right) - Q\left(g\left(\mathcal{E}\right),\mathbf{X}_U\right)\right\vert \geq K_1 n^{-\min\left\{-1/u,-1/2\right\}}\log\left(n\right)^{u+\beta + 1} + t \right) \leq \\
\PP&\Big(\left\vert Q_n\left(g\left(\hat{\boldsymbol{\epsilon}}\right),\mathbf{x}_U\right) - Q_n\left(g\left(\boldsymbol{\epsilon}\right),\mathbf{x}_U\right)\right\vert + \left\vert Q_n\left(g\left(\boldsymbol{\epsilon}\right),\mathbf{x}_U\right) - Q\left(g\left(\mathcal{E}\right),\mathbf{X}_U\right)\right\vert \geq \\
& K_1 n^{-\min\left\{-1/u,-1/2\right\}}\log\left(n\right)^{u+\beta + 1} + t \Big) \leq \\
\PP& \left(\left\vert Q_n\left(g\left(\hat{\boldsymbol{\epsilon}}\right),\mathbf{x}_U\right) - Q_n\left(g\left(\boldsymbol{\epsilon}\right),\mathbf{x}_U\right)\right\vert \geq \dfrac{t}{2} \right)+\\
\PP&\left(\left\vert Q_n\left(g\left(\boldsymbol{\epsilon}\right),\mathbf{x}_U\right) - Q\left(g\left(\mathcal{E}\right),\mathbf{X}_U\right)\right\vert \geq K_1 n^{-\min\left\{-1/u,-1/2\right\}}\log\left(n\right)^{u+\beta + 1} + \dfrac{t}{2} \right) \leq \\
& \dfrac{K_2\left(n\right)}{t}+K_3 \exp\left(-K_4 nt^2\right) \leq K_5 \max\left\{\dfrac{D\left(n\right)}{t}, \exp\left(-K_4 nt^2\right)\right\},
\end{align*}
where we used Lemma 14.2 in \citep{azadkia2021simple} in the second to last inequality.
Finally, we can follow the proof idea of Lemma 16.3 in \citep{azadkia2021simple} with the given probability bound showing that the probability of $\hat{S}$ being insufficient goes to $0$.


\subsection{Proof of Proposition \ref{prop:discrete}}\label{app:proof-discrete}
For the least squares parameter, we have
\begin{equation*}
\hat{\beta}=\dfrac{\mathbf{x}^\top \mathbf{y}}{\mathbf{x}^\top \mathbf{x}}= \beta + \dfrac{\mathbf{x}^\top \boldsymbol{\epsilon}}{\mathbf{x}^\top \mathbf{x}}, \qquad \PP\left(\hat{\beta} = \beta \right) = \PP\left(\mathbf{x}^\top \boldsymbol{\epsilon} = 0 \right) = 0
\end{equation*}
since $X$ is a continuous random variable. However, for large enough sample size, it holds (with high probability)
\begin{equation*}
i \in \underset{l}{\text{arg min}} \left\vert \hat{\epsilon}_i - \epsilon_l\right\vert \ \forall i,
\end{equation*}
i.e., the estimated residuals scatter closely around the true value from the discrete set. There are roughly $n/k$ observations per possible value of $\mathcal{E}$, and, due to the linear dependence, around each value, the ordering of $\hat{\boldsymbol{\epsilon}}$ corresponds to the ordering of $\mathbf{x}$ or is exactly inverted. Therefore,
\begin{equation*}
\hat{R}_i \mod \dfrac{n}{k} \approx \hat{R}_{M\left(i\right)} \mod  \dfrac{n}{k} \ \text{and} \ \hat{R}_{M\left(i\right)}\vert \hat{R}_i \dot\sim \hat{R}_i \mod \dfrac{n}{k} + \dfrac{n}{k} \text{Unif}\left\{0, \ldots, k-1\right\}.
\end{equation*}
Since the $\hat{\epsilon}_i$ all have distinct values, it holds
\begin{align*}
\sum_{i=1}^n \hat{L}_i & = \sum_{i=1}^n i = \dfrac{n^2 + n}{2} \ \text{and} \sum_{i=1}^n \hat{L}_i^2 = \sum_{i=1}^n i^2 = \dfrac{n\left(n + 1\right)\left(2n + 1\right)}{6} \ \text{such that} \\
T_n\left(\hat{\boldsymbol{\epsilon}},\mathbf{x}\right) & = \dfrac{n \sum_{i=1}^n \min\left\{\hat{R}_i, \hat{R}_{M \left(i\right)}\right\} - \dfrac{n\left(n + 1\right)\left(2n + 1\right)}{6}}{\dfrac{n^3 + n^2}{2}-\dfrac{n\left(n + 1\right)\left(2n +1 \right)}{6}}.
\end{align*}
We consider the only random term
\begin{align*}
\EE\left[\sum_{i=1}^n \min\left\{\hat{R}_i, \hat{R}_{M \left(i\right)}\right\}\right] & = \EE\left[\sum_{\hat{R}_i=1}^n \min\left\{\hat{R}_i, \hat{R}_{M \left(i\right)}\right\}\right] = \EE\left[\sum_{\hat{R}_i=1}^n \min\left\{\hat{R}_i, \hat{R}_{M \left(i\right)}\right\}\vert \hat{R}_1, \ldots, \hat{R}_n\right] \\
 & = \sum_{\hat{R}_i=1}^n \EE\left[\min\left\{\hat{R}_i, \hat{R}_{M \left(i\right)}\right\}\vert \hat{R}_1, \ldots, \hat{R}_n\right] = \sum_{\hat{R}_i=1}^n \EE\left[\min\left\{\hat{R}_i, \hat{R}_{M \left(i\right)}\right\}\vert \hat{R}_i\right].
\end{align*}
The first equality holds as summing over all $i$ is the same as summing over all ranks. As the problem is permutation invariant, conditioning on all ranks does not change the expectation. Under the conditioning, the ranks are deterministic and linearity of expectation applies. Finally, knowing any rank apart from $\hat{R}_i$ does not influence $\min\left\{\hat{R}_i, \hat{R}_{M \left(i\right)}\right\}$. We analyse the expectation under the approximate conditional distribution as given above. If $\hat{R}_i \leq n/k$, $\min\left\{\hat{R}_i, \hat{R}_{M \left(i\right)}\right\} = \hat{R}_i$. If $n/k < \hat{R}_i \leq 2n/k$ and the uniformly chosen number is $0$, $\min\left\{\hat{R}_i,\hat{R}_{M \left(i\right)}\right\} = \hat{R}_i -n/k$. This has probability $1/k$. Otherwise, $\min\left\{\hat{R}_i,\hat{R}_{M \left(i\right)}\right\} = \hat{R}_i$. Analogously, if $ln/k < \hat{R}_i \leq (l+1)n/k$ for some integer $0\leq l < k$, i.e., $l = \max\left\{r \in \mathbb{N}_0\vert r < \hat{R}_i k/n\right\}$, it holds under the approximate distribution
\begin{equation*}
\tilde{\EE}\left[\min\left\{\hat{R}_i, \hat{R}_{M \left(i\right)}\right\}\vert \hat{R}_i\right] = \hat{R}_i - \dfrac{1}{k} \sum_{r=0}^l \dfrac{rn}{k} = \hat{R}_i - \dfrac{n}{2k^2}\left(l^2 + l\right).
\end{equation*}
For each possible value of $l$, there are $n/k$ ranks such that $ln/k < \hat{R}_i \leq (l+1)n/k$. Therefore,
\begin{align*}
&\sum_{\hat{R}_i=1}^n \tilde{\EE}\left[\min\left\{\hat{R}_i, \hat{R}_{M \left(i\right)}\right\}\vert \hat{R}_i\right] =\sum_{\hat{R}_i=1}^n \hat{R}_i - \dfrac{n^2}{2k^3}\sum_{l=0}^{k-1}\left(l^2 + l\right)= \\
& \dfrac{n^2 + n}{2}-\dfrac{n^2}{2k^3}\left(\dfrac{k\left(k-1\right)\left(2k -1\right)}{6}+ \dfrac{k\left(k-1\right)}{2}\right) = \dfrac{n^2 + n}{2}-\dfrac{n^2}{6 k^3}\left(k^3 - k \right)
\end{align*}
Then,
\begin{align*}
\EE\left[T_n\left(\hat{\boldsymbol{\epsilon}},\mathbf{x}\right)\right] \approx \dfrac{\dfrac{n^3 + n^2}{2}-\dfrac{n\left(n + 1\right)\left(2n +1 \right)}{6}-\dfrac{n^3}{6 k^3}\left(k^3 - k \right)}{\dfrac{n^3 + n^2}{2}-\dfrac{n\left(n + 1\right)\left(2n +1 \right)}{6}} \overset{n \rightarrow \infty}{\rightarrow} \dfrac{\dfrac{1}{6}-\dfrac{1}{6} + \dfrac{1}{6k^2}}{\dfrac{1}{6}} = \dfrac{1}{k^2}
\end{align*}
To make the proof complete the proof, we need to show that
\begin{equation*}
\left\vert \sum_{\hat{R}_i=1}^n \tilde{\EE}\left[\min\left\{\hat{R}_i, \hat{R}_{M \left(i\right)}\right\}\vert \hat{R}_i\right] -\EE\left[\min\left\{\hat{R}_i, \hat{R}_{M \left(i\right)}\right\}\vert \hat{R}_i\right]\right\vert = {\scriptstyle \mathcal{O}} \left(n^2\right).
\end{equation*}
We even control
\begin{equation*}
\sum_{\hat{R}_i=1}^n \left\vert \tilde{\EE}\left[\min\left\{\hat{R}_i, \hat{R}_{M \left(i\right)}\right\}\vert \hat{R}_i\right] -\EE\left[\min\left\{\hat{R}_i, \hat{R}_{M \left(i\right)}\right\}\vert \hat{R}_i\right]\right\vert.
\end{equation*}
For arbitrary conditioning events $A$, we have
\begin{align*}
\sum_{\hat{R}_i=1}^n &\left\vert \tilde{\EE}\left[\min\left\{\hat{R}_i, \hat{R}_{M \left(i\right)}\right\}\vert \hat{R}_i\right] -\EE\left[\min\left\{\hat{R}_i, \hat{R}_{M \left(i\right)}\right\}\vert \hat{R}_i\right]\right\vert =\\
\sum_{\hat{R}_i=1}^n &\Big\vert \tilde{\EE}\left[\min\left\{\hat{R}_i, \hat{R}_{M \left(i\right)}\right\}\vert \hat{R}_i\right] -\EE\left[\min\left\{\hat{R}_i, \hat{R}_{M \left(i\right)}\right\}\vert \hat{R}_i, A\right]\PP\left(A\right)-\\
&\phantom{\Big\vert}\EE\left[\min\left\{\hat{R}_i, \hat{R}_{M \left(i\right)}\right\}\vert \hat{R}_i, A^c \right]\PP\left(A^c\right)\Big\vert \leq \\
\sum_{\hat{R}_i=1}^n &\left\vert \tilde{\EE}\left[\min\left\{\hat{R}_i, \hat{R}_{M \left(i\right)}\right\}\vert \hat{R}_i\right] -\EE\left[\min\left\{\hat{R}_i, \hat{R}_{M \left(i\right)}\right\}\vert \hat{R}_i, A\right]\right\vert \PP\left(A\right)+\\
\sum_{\hat{R}_i=1}^n &\left\vert \tilde{\EE}\left[\min\left\{\hat{R}_i, \hat{R}_{M \left(i\right)}\right\}\vert \hat{R}_i\right] -\EE\left[\min\left\{\hat{R}_i, \hat{R}_{M \left(i\right)}\right\}\vert \hat{R}_i, A^c\right]\right\vert \PP\left(A^c\right) \leq \\
\sum_{\hat{R}_i=1}^n & n \PP\left(A\right)+ \sum_{\hat{R}_i=1}^n \left\vert \tilde{\EE}\left[\min\left\{\hat{R}_i, \hat{R}_{M \left(i\right)}\right\}\vert \hat{R}_i\right] -\EE\left[\min\left\{\hat{R}_i, \hat{R}_{M \left(i\right)}\right\}\vert \hat{R}_i, A^c\right]\right\vert = \\
\sum_{\hat{R}_i=1}^n &\left\vert \tilde{\EE}\left[\min\left\{\hat{R}_i, \hat{R}_{M \left(i\right)}\right\}\vert \hat{R}_i\right] -\EE\left[\min\left\{\hat{R}_i, \hat{R}_{M \left(i\right)}\right\}\vert \hat{R}_i, A^c\right]\right\vert + n^2 \PP\left(A\right).
\end{align*}
Hence, we can ignore events with vanishing probability.
Let $v_1, \ldots, v_k$ be the attainable values of $\mathcal{E}$ and
\begin{equation*}
n_t = \sum_{i=1}^n \mathbbm{1}_{\left\{\epsilon_i = v_t\right\}} \sim \text{Binom}\left(n, \dfrac{1}{k}\right). 
\end{equation*}
Define the event 
\begin{equation*}
A = \left\{\underset{t}{\max}\left\vert n_t - n/k \right\vert > n^{3/4}\cup \exists i : \ i \not\in \underset{l}{\text{arg min}} \left\vert \hat{\epsilon}_i - \epsilon_l\right\vert\right\}.
\end{equation*}
By the Markov inequality and a union bound, this event has vanishing probability, so we must only control
\begin{align*}
&\sum_{\hat{R}_i=1}^n\left\vert\tilde{\EE}\left[\min\left\{\hat{R}_i, \hat{R}_{M \left(i\right)}\right\}\vert \hat{R}_i\right] -\EE\left[\min\left\{\hat{R}_i, \hat{R}_{M \left(i\right)}\right\}\vert \hat{R}_i, A^c\right]\right\vert.
\end{align*}
Under $A^c$, there are only $\mathcal{O}\left(n^{3/4}\right)$ ranks $\hat{R}_i$ for which $\epsilon_i \neq v_{l+1}$ with $l = \max\left\{r \in \mathbb{N}_0\vert r < \hat{R}_i k/n\right\}$ is possible. Summing over these leads to another ${\scriptstyle \mathcal{O}} \left(n^2\right)$ term and can be ignored. Consider the $\hat{R}_i$ for which $A_i \coloneqq \left\{\epsilon_i = v_{l+1}\right\}$ holds.  Assume without loss of generality that $\hat{\beta} < \beta$ such that larger $x_i$ leads to larger $\hat{\epsilon}_i$. Let $F_X\left(\cdot\right)$ be the cumulative distribution function of $X$. For given $n_1, \ldots, n_k$, we have
\begin{equation*}
x_i = F^{-1}_X\left(\dfrac{\hat{R}_i -\sum_{r=1}^l{n_r}}{n_{l+1}}\right)+\mathcal{O}_p\left(n^{-1/2}\right).
\end{equation*}
Thus, one can condition on $x_i$ being in a $n^{-1/4}$ range around the theoretical quantile for any $n_1, \ldots, n_k$ fulfilling $A_i$. Call this event, whose complementary event has vanishing probability, $B_i$. It remains to control
\begin{align*}
&\EE\left[\min\left\{\hat{R}_i, \hat{R}_{M \left(i\right)}\right\}\vert \hat{R}_i,A^c, A_i, B_i \right]=\\\sum_{r=1}^k &\EE\left[\min\left\{\hat{R}_i, \hat{R}_{M \left(i\right)}\right\}\vert \hat{R}_i,A^c, A_i, B_i, \epsilon_{M\left(i\right)}=v_r \right]\PP\left(\epsilon_{M\left(i\right)}=v_r \vert\hat{R}_i,A^c, A_i, B_i\right)=\\
\sum_{r=1}^k &\EE\left[\min\left\{\hat{R}_i, \hat{R}_{M \left(i\right)}\right\}\vert \hat{R}_i, A^c, A_i, B_i, \epsilon_{M\left(i\right)}=v_r \right]\left(\PP\left(\epsilon_{M\left(i\right)}=v_r \vert \hat{R}_i\right)+ {\scriptstyle \mathcal{O}}\left(1\right)\right)=\\
\sum_{r=1}^k &\EE\left[\min\left\{\hat{R}_i, \hat{R}_{M \left(i\right)}\right\}\vert \hat{R}_i, A^c, A_i, B_i, \epsilon_{M\left(i\right)}=v_r \right]\left(\dfrac{1}{k} + {\scriptstyle \mathcal{O}}\left(1\right)\right).
\end{align*}
If $\epsilon_{M\left(i\right)}>\epsilon_i$, it holds $\min\left\{\hat{R}_i, \hat{R}_{M \left(i\right)}\right\} = \hat{R}_i$ and we get the right contribution. If $\epsilon_{M\left(i\right)}=\epsilon_i$, the conditional expectation is in $\left[\hat{R}_i-1, \hat{R}_i\right]$, i.e., only a $\mathcal{O}\left(1\right)$ deviation. If $v_{m+1}=\epsilon_{M\left(i\right)}<\epsilon_i$ 
\begin{align*}
\min\left\{\hat{R}_i, \hat{R}_{M \left(i\right)}\right\} & = \hat{R}_{M \left(i\right)}=\sum_{r=1}^m n_r + \sum_{l: \epsilon_l = v_{m+1}} \mathbbm{1}_{\left\{x_l \leq x_{M\left(i\right)}\right\}}\\
&=\sum_{r=1}^m n_r + \sum_{l: \epsilon_l = v_{m+1}} \mathbbm{1}_{\left\{x_l \leq x_i\right\}} + \mathbbm{1}_{\left\{x_{M\left(i\right)} > x_i\right\}}.
\end{align*}
Under the given conditioning, this is
\begin{align*}
&\sum_{r=1}^m n_r + n_{m+1} \left(\dfrac{\hat{R}_i -\sum_{r=1}^l{n_r}}{n_{l+1}}+{\scriptstyle \mathcal{O}}\left(1\right)\right)+\mathcal{O}\left(1\right)=\\
&\dfrac{nm}{k}+ \mathcal{O}\left(n^{3/4}\right)+\left(\dfrac{n}{k}+\mathcal{O}\left(n^{3/4}\right)\right)\dfrac{\hat{R}_i-nl/k + \mathcal{O}\left(n^{3/4}\right)}{n/k + \mathcal{O}\left(n^{3/4}\right)}+{\scriptstyle \mathcal{O}}\left(n\right)=\\
&\dfrac{nm}{k} + \left(1 + \mathcal{O}\left(n^{-1/4}\right)\right)\left(\hat{R}_i - \dfrac{nl}{k} + \mathcal{O}\left(n^{3/4}\right)\right)+{\scriptstyle \mathcal{O}}\left(n\right)=\dfrac{nm}{k} +\hat{R}_i - \dfrac{nl}{k} +\mathcal{O}\left(n^{3/4}\right)+{\scriptstyle \mathcal{O}}\left(n\right)=\\
&\hat{R}_i - \dfrac{n\left(l-m\right)}{k} + {\scriptstyle \mathcal{O}}\left(n\right).
\end{align*}
In summary
\begin{equation*}
 \EE\left[\min\left\{\hat{R}_i, \hat{R}_{M \left(i\right)}\right\}\vert \hat{R}_i, A^c, A_i, B_i \right] = \hat{R}_i - \dfrac{1}{k}\sum_{m=0}^{l-1}\dfrac{n\left(l-m\right)}{k} + {\scriptstyle \mathcal{O}}\left(n\right)= \hat{R}_i - \dfrac{1}{k}\sum_{r=0}^{l}\dfrac{r}{k} + {\scriptstyle \mathcal{O}}\left(n\right)
\end{equation*}
Therefore, each term deviates with ${\scriptstyle \mathcal{O}}\left(n\right)$ from the approximate expectation. Summing over $\mathcal{O}\left(n\right)$ such deviations leads to ${\scriptstyle \mathcal{O}}\left(n^2\right)$ as desired.

\subsection{Proof of Theorem \ref{theo:H0j-het}}
Due to \ref{ass:add-mult-sep}, we have
\begin{align*}
\EE\left[Y \vert \mathbf{X}\right] =& f_{\mathbf{X}_U Y}\left(\mathbf{X}_U, \mathbf{X}_{\text{PA}\left(Y\right) \setminus U}\right) + g_{\mathbf{X}_U Y}\left(\mathbf{X}_U, \mathbf{X}_{\text{PA}\left(Y\right) \setminus U}\right) \EE\left[f_{\mathbf{H}Y}\left(\mathbf{H}_{\text{PA}\left(Y\right)}, \mathbf{X}_{\text{PA}\left(Y\right) \setminus U}\right) \vert \mathbf{X}\right] \\
\text{Var}\left(Y \vert \mathbf{X}\right) =& g^2_{\mathbf{X}_U Y}\left(\mathbf{X}_U, \mathbf{X}_{\text{PA}\left(Y\right) \setminus U}\right)\\
&\left(\EE\left[f^2_{\mathbf{H}Y}\left(\mathbf{H}_{\text{PA}\left(Y\right)}, \mathbf{X}_{\text{PA}\left(Y\right) \setminus U}\right) \vert \mathbf{X}\right]-\EE\left[f_{\mathbf{H}Y}\left(\mathbf{H}_{\text{PA}\left(Y\right)}, \mathbf{X}_{\text{PA}\left(Y\right) \setminus U}\right) \vert \mathbf{X}\right]^2\right)\\
\mathcal{E} =&\dfrac{f_{\mathbf{H}Y}\left(\mathbf{H}_{\text{PA}\left(Y\right)}, \mathbf{X}_{\text{PA}\left(Y\right) \setminus U}\right) - \EE\left[f_{\mathbf{H}Y}\left(\mathbf{H}_{\text{PA}\left(Y\right)}, \mathbf{X}_{\text{PA}\left(Y\right) \setminus U}\right) \vert \mathbf{X}\right]}{\sqrt{\EE\left[f^2_{\mathbf{H}Y}\left(\mathbf{H}_{\text{PA}\left(Y\right)}, \mathbf{X}_{\text{PA}\left(Y\right) \setminus U}\right) \vert \mathbf{X}\right]-\EE\left[f_{\mathbf{H}Y}\left(\mathbf{H}_{\text{PA}\left(Y\right)}, \mathbf{X}_{\text{PA}\left(Y\right) \setminus U}\right) \vert \mathbf{X}\right]^2}}
\end{align*}
As in Section \ref{proof:H0j}
\begin{align*}
f_{\mathbf{H}Y}\left(\mathbf{H}_{\text{PA}\left(Y\right)}, \mathbf{X}_{\text{PA}\left(Y\right)\setminus j}\right)  \perp \mathbf{X}_U \vert \mathbf{X}_{-U} \quad \text{and} \quad
\EE\left[f_{\mathbf{H}Y}\left(\mathbf{H}_{\text{PA}\left(Y\right)}, \mathbf{X}_{\text{PA}\left(Y\right) \setminus U}\right) \vert \mathbf{X}\right] = & \perp \mathbf{X}_U \vert \mathbf{X}_{-U} \\
\text{accordingly} \quad \EE\left[f^2_{\mathbf{H}Y}\left(\mathbf{H}_{\text{PA}\left(Y\right)}, \mathbf{X}_{\text{PA}\left(Y\right) \setminus U}\right) \vert \mathbf{X}\right] \perp \mathbf{X}_U \vert \mathbf{X}_{-U}\quad \text{such that} \quad \mathcal{E} & \perp \mathbf{X}_U  \vert \mathbf{X}_{-U}.
\end{align*}
\ref{ass:markov} together with \ref{ass:indep} implies that only terms involving $\mathbf{X}_U$ can be different for the counterfactual; see Section \ref{proof:H0j}. In particular, $\mathcal{E}$ cannot change. Hence, $Y$ changes from
\begin{align*}
y & = \EE\left[Y\vert \mathbf{X} = \mathbf{x}\right] + \sqrt{\text{Var}\left(Y\vert \mathbf{X}=\mathbf{x}\right)}  \epsilon \quad \text{to} \\
y' & = \EE\left[Y \vert \mathbf{X}_U = \mathbf{x}'_U, \mathbf{X}_{-U} = \mathbf{x}_{-U}\right] + \sqrt{\text{Var}\left(Y \vert \mathbf{X}_U = \mathbf{x}'_U, \mathbf{X}_{-U} = \mathbf{x}_{-U}\right)}  \epsilon
\end{align*}
which is as stated in the theorem.
\subsection{Proof of Theorem \ref{theo:heto-cons}}
We have the following supporting result.
\begin{lemm}\label{lemm:heto-conv}
Suppose that \ref{ass:f1} - \ref{ass:var0} hold. Then
\begin{equation*}
\left \vert \hat{\boldsymbol{\epsilon}}_i - \boldsymbol{\epsilon}_i \right \vert = {\scriptstyle \mathcal{O}}_p \left(1\right)
\end{equation*}
\end{lemm}
With Lemma \ref{lemm:heto-conv} we have replaced Assumption \ref{ass:suitable} which is the only missing part to reconstruct the asymptotic results as in Theorems \ref{theo:cons} and \ref{theo:cons2}.
\subsubsection{Proof of Lemma \ref{lemm:heto-conv}}
Let 
\begin{equation*}
V\left(\mathbf{x}_i\right) = f_2\left(\mathbf{x}_i\right) - f_1^2\left(\mathbf{x}_i\right) \quad \text{and} \quad \hat{V}\left(\mathbf{x}_i\right) = \hat{f}_2\left(\mathbf{x}_i\right) - \hat{f}_1^2\left(\mathbf{x}_i\right)
\end{equation*}
Note that
\begin{equation*}
\PP\left(\dfrac{1}{V\left(\mathbf{x}_i\right)} < \infty \right) = \PP\left(V\left(\mathbf{x}_i\right) > 0\right)=1 \quad \text{hence} \quad \dfrac{1}{V\left(\mathbf{x}_i\right)}=\mathcal{O}_p\left(1\right)\quad \text{likewise} \quad \dfrac{1}{\sqrt{V\left(\mathbf{x}_i\right)}}=\mathcal{O}_p\left(1\right).
\end{equation*}
Consider the difference
\begin{align*}
\left\vert V\left(\mathbf{x}_i\right)- \hat{V}\left(\mathbf{x}_i\right) \right\vert & = \left\vert f_2\left(\mathbf{x}_i\right) - \left(f_2\left(\mathbf{x}_i\right) + {\scriptstyle \mathcal{O}}_p \left(1\right)\right) - f_1^2\left(\mathbf{x}_i\right) + \left( f_1\left(\mathbf{x}_i\right) + {\scriptstyle \mathcal{O}}_p \left(1\right)\right)^2 \right\vert \\
& \leq {\scriptstyle \mathcal{O}}_p \left(1\right) + \left\vert f_1\left(\mathbf{x}_i\right) \right \vert {\scriptstyle \mathcal{O}}_p \left(1\right) = {\scriptstyle \mathcal{O}}_p \left(1\right).
\end{align*}
In the last equality we use $\EE\left[f_1\left(\mathbf{x}_i\right)\right] = \EE\left[y_i\right] < \infty$, otherwise regression would not be possible. Hence,
\begin{equation*}
\underset{n \rightarrow \infty}{\lim}\PP \left(\hat{V}\left(\mathbf{x}_i\right) \leq 0\right) \leq \underset{n \rightarrow \infty}{\lim}\PP \left( \left\vert V\left(\mathbf{x}_i\right)- \hat{V}\left(\mathbf{x}_i\right) \right\vert \geq V\left(\mathbf{x}_i\right)\right) =\underset{n \rightarrow \infty}{\lim}\PP \left(\dfrac{\left\vert V\left(\mathbf{x}_i\right)- \hat{V}\left(\mathbf{x}_i\right) \right\vert}{V\left(\mathbf{x}_i\right)} \geq 1\right)=0.
\end{equation*}
Therefore, we will forthcoming condition on $\hat{V}\left(\mathbf{x}_i\right)$ being positive which is asymptotically negligible. We now compare the standard deviation and its estimate and consider the event that the difference is either large or not defined. Fix some $\eta > 0$.
\begin{align*}
\underset{n \rightarrow \infty}{\lim}& \PP \left(\left \vert\sqrt{V\left(\mathbf{x}_i\right)} - \sqrt{\hat{V}\left(\mathbf{x}_i\right)}\right \vert \geq \eta \cup \hat{V}\left(\mathbf{x}_i\right) < 0\right)\\
 = \underset{n \rightarrow \infty}{\lim} &\PP \left(\left \vert\sqrt{V\left(\mathbf{x}_i\right)} - \sqrt{\hat{V}\left(\mathbf{x}_i\right)}\right \vert \geq \eta \cup \hat{V}\left(\mathbf{x}_i\right) < 0  \vert \hat{V}\left(\mathbf{x}_i\right) > 0\right)\PP\left(\hat{V}\left(\mathbf{x}_i\right) > 0\right) +\\
 & \PP \left(\left \vert\sqrt{V\left(\mathbf{x}_i\right)} - \sqrt{\hat{V}\left(\mathbf{x}_i\right)}\right \vert \geq \eta \cup \hat{V}\left(\mathbf{x}_i\right) < 0  \vert \hat{V}\left(\mathbf{x}_i\right) \leq 0\right)\PP\left(\hat{V}\left(\mathbf{x}_i\right) \leq 0\right)\\
\leq \underset{n \rightarrow \infty}{\lim} &\PP \left(\left \vert\sqrt{V\left(\mathbf{x}_i\right)} - \sqrt{\hat{V}\left(\mathbf{x}_i\right)}\right \vert \geq \eta \cup \hat{V}\left(\mathbf{x}_i\right) < 0  \vert \hat{V}\left(\mathbf{x}_i\right) > 0\right) +  \PP\left(\hat{V}\left(\mathbf{x}_i\right) \leq 0\right)\\
=\underset{n \rightarrow \infty}{\lim} &\PP \left(\left \vert\sqrt{V\left(\mathbf{x}_i\right)} - \sqrt{\hat{V}\left(\mathbf{x}_i\right)}\right \vert \geq \eta \vert \hat{V}\left(\mathbf{x}_i\right) > 0\right)=\underset{n \rightarrow \infty}{\lim} \PP \left(\left \vert\dfrac{V\left(\mathbf{x}_i\right)- \hat{V}\left(\mathbf{x}_i\right)}{\sqrt{V\left(\mathbf{x}_i\right)} + \sqrt{\hat{V}\left(\mathbf{x}_i\right)}}\right \vert \geq \eta \vert \hat{V}\left(\mathbf{x}_i\right) > 0\right)\\
\leq \underset{n \rightarrow \infty}{\lim} &\PP \left(\left \vert\dfrac{V\left(\mathbf{x}_i\right)- \hat{V}\left(\mathbf{x}_i\right)}{\sqrt{V\left(\mathbf{x}_i\right)}}\right \vert \geq \eta \vert \hat{V}\left(\mathbf{x}_i\right) > 0\right)\leq \underset{n \rightarrow \infty}{\lim} \PP \left(\left \vert\dfrac{V\left(\mathbf{x}_i\right)- \hat{V}\left(\mathbf{x}_i\right)}{\sqrt{V\left(\mathbf{x}_i\right)}}\right \vert \geq \eta\right) / \PP \left(\hat{V}\left(\mathbf{x}_i\right) > 0\right)=0.
\end{align*}
Consider the residuals assuming a positive variance estimate.
\begin{align*}
\left \vert \hat{\boldsymbol{\epsilon}}_i - \boldsymbol{\epsilon}_i \right \vert & = \left \vert \dfrac{y_i - \hat{f}_i\left(\mathbf{x}_i\right)}{\sqrt{\hat{V}\left(\mathbf{x}_i\right)}} - \dfrac{y_i - f_i\left(\mathbf{x}_i\right)}{\sqrt{V\left(\mathbf{x}_i\right)}} \right \vert = \left\vert  \dfrac{\left(y_i - \hat{f}_i\left(\mathbf{x}_i\right) \right)\sqrt{V\left(\mathbf{x}_i\right)} - \left(y_i - f_i\left(\mathbf{x}_i\right)\right)\sqrt{\hat{V}\left(\mathbf{x}_i\right)}}{\sqrt{\hat{V}\left(\mathbf{x}_i\right)}\sqrt{V\left(\mathbf{x}_i\right)} } \right\vert  \\
& =\left\vert  \dfrac{\left(f_i\left(\mathbf{x}_i\right) - \hat{f}_i\left(\mathbf{x}_i\right) \right)\sqrt{V\left(\mathbf{x}_i\right)} + \left(y_i - f_i\left(\mathbf{x}_i\right)\right)\left(\sqrt{V\left(\mathbf{x}_i\right)} - \sqrt{\hat{V}\left(\mathbf{x}_i\right)}\right)}{\sqrt{\hat{V}\left(\mathbf{x}_i\right)}\sqrt{V\left(\mathbf{x}_i\right)} } \right\vert \\
\leq & \dfrac{1}{\sqrt{\hat{V}\left(\mathbf{x}_i\right)}}\left(\left \vert f_i\left(\mathbf{x}_i\right) - \hat{f}_i\left(\mathbf{x}_i\right)\right\vert + \left\vert \epsilon_i\right\vert \left\vert \sqrt{V\left(\mathbf{x}_i\right)} - \sqrt{\hat{V}\left(\mathbf{x}_i\right)} \right\vert\right).
\end{align*}
Arguing similarly as before, we have
\begin{equation*}
\underset{n \rightarrow \infty}{\lim}\PP \left(\left \vert \hat{\boldsymbol{\epsilon}}_i - \boldsymbol{\epsilon}_i \right \vert \geq \eta \cup \hat{V}\left(\mathbf{x}_i\right) < 0\right) = 0 .
\end{equation*}
If we replace $\hat{\epsilon}_i$ by an arbitrary value in case of a non-positive variance estimate, it holds 
\begin{equation*}
\underset{n \rightarrow \infty}{\lim}\PP \left(\left \vert \hat{\boldsymbol{\epsilon}}_i - \boldsymbol{\epsilon}_i \right \vert \geq \eta\right) = 0 .
\end{equation*}
\end{document}